\documentclass[a4paper,11pt]{article}
\pdfoutput=1 % if your are submitting a pdflatex (i.e. if you have
\usepackage{jheppub} % for details on the use of the package, please
\usepackage{xcolor}
\usepackage{pagecolor,lipsum}% http://ctan.org/pkg/{pagecolor,lipsum}
\usepackage{mdframed}
\usepackage{verbatim}
\usepackage{subcaption,graphicx}
\usepackage[T1]{fontenc} % if needed
\usepackage{mathrsfs}
\usepackage{arydshln}
\usepackage{color}
\usepackage{slashed}
\usepackage{epsfig}
\usepackage{enumerate}
\usepackage{color}
\usepackage{wrapfig}
\usepackage{amsfonts}
\usepackage{multirow}
\usepackage{amsthm}
\usepackage{amsmath}
\usepackage{tikz}
\usepackage{float}
\usepackage[numbers]{natbib}
\usepackage{blindtext}
\usepackage{multicol}
\usepackage{color}
\usepackage{pdflscape}
 \usepackage[normalem]{ulem}

\newcommand{\ba}{  \begin{array}}
\newcommand{\ea}{\end{array}}

\newcommand{\bd}{  \begin{displaymath}}
\newcommand{\ed}{\end{displaymath}}

\def\l{\lambda}
 
 \def\mhpm{m_{H^{\pm}}}

\newcommand{\bsube}{  \begin{subequation}}
\newcommand{\esube}{\end{subequation}}
\newcommand{ \bea}{  \begin{eqnarray}}
\newcommand{ \eea}{\end{eqnarray}}
\newcommand{\bal}{  \begin{align}}
\newcommand{\ealign}{\end{align}}
\newcommand{\eal}{\end{align}}
\newcommand{ \bean}{  \begin{enumerate}}
\newcommand{ \eean}{\end{enumerate}}
\newcommand{\nn}{\nonumber}

\newcommand{\ra}{\rightarrow}

\newcommand{\gsim}{\:\raisebox{-0.5ex}{$\stackrel{\textstyle>}{\sim}$}\:}

 \usepackage[normalem]{ulem}
 
 \definecolor{darkgreen}{cmyk}{1,0,1,0.4}
 \definecolor{pink}{cmyk}{0.4,1,0.3,0}

\title{Leptonic $g-2$ anomaly in an extended Higgs sector with vector-like leptons}
\author[a,\,\dagger]{Hrishabh Bharadwaj,}
\author[b,\,\#]{Sukanta Dutta,}
\author[a,\,\$]{Ashok Goyal,}

\affiliation[a]{Department of Physics $\&$ Astrophysics, University of Delhi, Delhi, India.}
\affiliation[b]{SGTB Khalsa College, University of Delhi, Delhi, India.}

\emailAdd{${}^\dagger$ Corresponding~Author: hrishabhphysics@gmail.com}
\emailAdd{${}^\#$Sukanta.Dutta@gmail.com}
\emailAdd{${}^{\$}$agoyal45@yahoo.com}

\abstract{We address the observed discrepancies in the anomalous magnetic  dipole  moments (MDM) of the muon and electron  by extending the inert two Higgs Doublet Model (2HDM) with  SM gauge singlet complex scalar field and  singlet Vector-like  Lepton (VLL)  field. We obtain the allowed parameter space constrained from the Higgs decays to gauge Bosons at LHC, LEP II data and electro-weak precision measurements. The muon and electron MDM’s are then explained within a common parameter space for different sets of allowed couplings and masses of the model particles.}
\keywords{Vector-like lepton, muon $g-2$, electron $g-2$, 2HDM }

\begin{document} 

\maketitle
\flushbottom
%%%%%%%%%%%%%%%%%%%%%%%%%%%%%%%%%%%%%%%%%%%%%%%%%%%%%%%%%%%%%%%%%%%%%%%%%%%%%%%%%%%%%%%%%
\section{Introduction}
The anomalous magnetic moment of the electron and muon has been measured to an unprecedented precision and its deviation with the theoretically calculated value in the Standard Model (SM) \cite{Keshavarzi:2018mgv,Blum:2018mom}  and it may as well be a portent of new physics beyond the SM. The estimated value of the anomalous MDM  of muon  \cite{Muong-2:2021ojo} 
\bea
a_{\mu}^{\rm FNAL} &=& 116592040(54) \times 10^{-11}
\eea
 from recent measurements by  {\it G-2 Collaboration} validates the earlier  observations from the Brookhaven National Laboratory  E821 experiment \cite{Bennett:2006fi,Brown:2001mga}. The combined measurements for $\mu^-$ and $\mu^+$ from both these experiments result in $a_\mu^{\rm expt} = 116 592 061(41) \times 10^{-11}$ \cite{Muong-2:2021ojo}. Comparing with the recent theoretical prediction  in SM $a_\mu^{\rm SM} = 116591810(43) \times 10^{-11}$ \cite{Aoyama:2020ynm}, a discrepancy of  $4.2 \sigma$ is observed and the deviation of anomalous MDM from SM prediction  is given as  \cite{Muong-2:2021ojo}
\bea
\Delta a_{\mu} &=&  \left(251 \pm 59 \right) \times 10^{-11}\label{delamu}
\eea

\par The principle uncertainty in the calculations of the SM contribution to $ a_\mu$  arises from the  hadronic  vacuum polarisation and from  light by light scattering contributions. Recently the Budapest-Marseille-Wuppertal collaboration  \cite{Borsanyi:2020mff} has computed the leading hadronic contribution to the muon anomalous MDM from lattice QCD and shown that there does not remain any discrepancy with the experiment. However, the HVP contribution has been estimated by  the authors of references \cite{Crivellin:2020zul,Keshavarzi:2020bfy,Colangelo:2020lcg}  indicating that this discrepancy far from being removed has only been shifted to the uncertainties in the $e^+e^-$ data and the electroweak fit. In the absence of more reliable and independently confirmed non- perturbative QCD  contribution, we will assume that BSM physics is indeed required to explain the discrepancy.

\par Another recent measurement of the fine structure constant $\alpha_{em}$ \cite{Parker:2018vye} has likewise resulted in a mild $\sim 2.4 \sigma$ discrepancy in experimental and theoretical prediction of the electron anomalous magnetic moment 
\bea
\Delta a_e&=& \left[-88\ \pm 28\, ({\rm expt.}) \pm 23\, (\alpha) \pm 2\,({\rm theory})\right] \times 10^{-14}\label{delae}
\eea

It is important to note that  anomalous MDM of the muon  is opposite in sign to that of an electron and is much larger in magnitude that can be accounted for, by the electron mass scaling $m_e^2/m_\mu^2$.

\par Various attempts for simultaneous explanation of the leptonic anomalous magnetic moment anomalies have been made in the past several years \cite{Hanneke:2008tm,Giudice:2012ms,Davoudiasl:2018fbb,Han:2018znu,Bauer:2019gfk,Hiller:2019mou,Cornella:2019uxs,Bigaran:2020jil,Jana:2020pxx,Calibbi:2020emz,Dutta:2020scq,Chen:2020tfr,Dorsner:2020aaz,Han:2015yys,Aad:2020xfq}. Models with axion-like particles (ALP) \cite{Bauer:2019gfk}, lepto-quarks \cite{CarcamoHernandez:2020pxw,Botella:2020xzf,ColuccioLeskow:2016dox,Crivellin:2020tsz}, vector-like leptons (VLL) \cite{Chun:2016hzs,Cherchiglia:2017uwv,Thomas:1998wy,Barman:2018jhz,Dermisek:2013gta,Falkowski:2013jya,Crivellin:2021rbq} and super-symmetric models \cite{Endo:2019bcj, Badziak:2019gaf, Liu:2018xkx, Abdullah:2019ofw, Arbelaez:2020rbq} have been employed with varying success to explain the anomaly. 
\par The two Higgs doublet model (2HDM) has been extensively employed in the literature to explain the muon magnetic moment anomaly \cite{Haba:2020gkr,Yang:2020bmh,Hati:2020fzp,Trodden:1998ym,Kim:1986ax,Gerard:2007kn,Broggio:2014mna,Cao:2009as,Ilisie:2015tra,Abe:2015oca}. The 2HDM model is the simplest extension of the SM. With an appropriate $Z_2$ symmetry, Type-X lepton specific 2HDM model with non-SM Higgs coupling to leptons being enhanced by $\tan\beta$, has been used to explain $(g-2)_\mu$. The solution, in general, requires large value of $\tan\beta$ and a light pseudo-scalar boson. The model is however strongly constrained by lepton precision observables and only a limited parameter space is available \cite{Cao:2009as,Gerard:2007kn,Abe:2015oca,Wang:2014sda,Chun:2015hsa}.
\par The 2HDM model has been extended with the inclusion of a real or complex singlet scalars with appropriate $Z_2$ symmetry to expand the available parameter space required to explain the muon magnetic moment anomaly \cite{Dutta:2018hcz,Dutta:2020scq}. Vector-like leptons have been introduced in the multi Higgs extension of the SM to relax the severe constraints discussed above. Inclusion of VLL in 2HDM enlarges the allowed parameter space consistent with the muon $g-2$ while still being within the theoretical and experimental bounds \cite{Chun:2016hzs,Cherchiglia:2017uwv,Thomas:1998wy,Barman:2018jhz,Dermisek:2013gta,Falkowski:2013jya}.
\par In the context of Lepton-portal Dark Matter models with the introduction of VLL or sleptons, an explanation of $(g-2)_\mu$ has been sought. It required simultaneous introduction of a VLL doublet and a singlet. In this model adherence to all experimental and theoretical bounds was found to be challenging \cite{Kawamura:2020qxo}.
\par A simultaneous explanation of the muon and electron $g-2$ anomaly was achieved in \cite{Hiller:2019mou} by introducing a VLL doublet and a singlet in the SM. The Higgs sector itself was extended by adding complex scalars in the TeV mass range. Similar study was done in \cite{Chun:2020uzw} where muon magnetic moment is obtained at the two-loop level with a sizable negative contribution to electron $g-2$ in the presence of vector-like leptons. A simultaneous explanation of the muon and electron $g-2$ anomaly in an inert lepton specific 2HDM model has been achieved in \cite{Han:2018znu}. In this model $Z_2$ symmetry is broken in the leptonic sector with non-universal Yukawa coupling between leptons and inert Higgs doublet. The model requires hierarchical couplings of inert Higgs doublet with leptons. Furthermore it was required that the Yukawa couplings for $\mu$ and $e/\tau$ leptons be opposite in sign and the parameter region was tightly constrained by Lepton flavor universality tests.
\par  Therefore, it is worthwhile to explore the variants of 2HDM which can  explain the anomalous MDMs of muon and electron. In section \ref{sec:model}, we formulate the viable model by augmenting  the  inert 2HDM  with a neutral complex scalar and a heavy vector-like charged lepton which are SM gauge singlets. We constrain the model parameters from Higgs decays, LEP data and precision measurements in section \ref{sec:EWconstraints}. We compute  the MDM of leptons in Section \ref{sec:MDM}  and conclude in Section \ref{sec:summary}.

%%%%%%%%%%%%%%%%%%%%%%%%%%%%%%
\section{The model}
\label{sec:model}
In order to simultaneously explain the muon and electron magnetic moment anomalies with common set of parameter values, we introduce a $Z_2$ symmetry in generic inert 2HDM which is allowed to be relaxed  in the leptonic sector with universal  Yukawa couplings. The lepton flavor universality (LFU) in the $\tau$ decays reported by the HFAG collaboration is then, trivially satisfied \cite{Amhis:2014hma}.
\par We begin the construction of the model by assigning the $Z_2$ parity quantum number for all the particle contents of the model.
\begin{table}[h!]\footnotesize
 \begin{center}
 \begin{tabular}{c|ccccccccccc}
 Fields &$Q_l$&$l_L$&$u_R$&$d_R$&$e_R$&$\Phi_1$&$\Phi_2$&$\Phi_3$&$\chi_L$&$\chi_R$&$V^\mu$\\\hline
 $SU(3)_c$&$3$&$1$&$3$&$3$&$1$&$1$&$1$&$1$&$1$&$1$&$G^\mu$\\
 $SU(2)_L$&$2$&$2$&$1$&$1$&$1$&$2$&$2$&$1$&$1$&$1$&$W^\mu_i$\\
 $U(1)_Y$&$\frac{1}{6}$&$-\frac{1}{2}$&$\frac{2}{3}$&$-\frac{1}{3}$&$-1$&$\frac{1}{2}$&$\frac{1}{2}$&$0$&$-1$&$-1$&$B^\mu$\\
 $Z_2$&$+$&$+$&$+$&$+$&$+$&$+$&$-$&$-$&$-$&$+$&$+$\\
  \end{tabular}
 \end{center}
\end{table}

\par Under $Z_2$ symmetry all the SM particles are assumed to be even whereas, scalar second doublet $\Phi_2$ and complex singlet $\Phi_3$ are odd. The left and the right chiral vector-like leptons are assumed to transform differently, namely, $\chi_L\to -\chi_L$ and $\chi_R\to \chi_R$. The $Z_2$ symmetry  ensures that the SM gauge bosons and fermions are forbidden to have direct interaction with the second (inert) Higgs doublet and additional complex scalar singlet.  We however, allow soft breaking of $Z_2$ symmetry by the vector-like lepton mass term and an explicit breaking of $Z_2$ symmetry in the Yukawa Lagrangian ${\cal L}_Y$ in order to facilitate coupling of SM leptons with $CP$ odd pseudo-scalars. 
\par   The Lagrangian is written as
\begin{subequations}
\bea
{\cal L}&=& {\cal L}_{\rm scalar}\ +\ {\cal L}_Y\ +\ {\cal L}_{\rm VL}
\eea
\bea
{\cal L}_{\rm scalar}&=& \left(D_\mu\Phi_1\right)^\dagger\ \left(D^\mu\Phi_1\right)\ +\ \left(D_\mu\Phi_2\right)^\dagger\ \left(D_\mu\Phi_2\right)\ +\ \left(D_\mu\Phi_3\right)^*\ \left(D_\mu\Phi_3\right)\ -\ V_{\rm scalar}\nn\\
\\
 V_{\rm scalar}&=& V_{\rm 2 HDM}\left(\Phi_1, \Phi_2\right)+ V_{\rm Singlet} \left(\Phi_3\right) +  V_{\rm Mix}\left(\Phi_1,\Phi_2, \Phi_3\right)\\
 &=& -\frac{1}{2} m_{11}^2  \left(\Phi_1^\dagger\Phi_1\right)  - \frac{1}{2} m_{22}^2  \left(\Phi_2^\dagger\Phi_2\right) + \frac{\lambda_1}{2} \left(\Phi_1^\dagger\Phi_1\right)^2 +\frac{\lambda_2}{2} \left(\Phi_2^\dagger\Phi_2\right)^2\nn\\
&&  + \lambda_3 \left(\Phi_1^\dagger\Phi_1\right)  \left(\Phi_2^\dagger\Phi_2\right) + \lambda_4 \left(\Phi_1^\dagger\Phi_2\right) \left(\Phi_2^\dagger\Phi_1\right) + \frac{1}{2} \left[\lambda_5\  \left(\Phi_1^\dagger\Phi_2\right)^2 +\ h.c. \right]\nn
\\
&&  -\frac{1}{2} m_{33}^2\ \Phi_3^*\Phi_3 + \frac{\lambda_8}{2} \left( \Phi_3^*\Phi_3 \right)^2   + \lambda_{11} \left\vert \Phi_1\right\vert^2 \Phi_3^*\Phi_3 + \lambda_{13} \left\vert\Phi_2\right\vert^2 \Phi_3^*\Phi_3
  \nn\\
  &&  -i\, \kappa\,\, \left[\left( \Phi_1^\dagger\Phi_2 + \Phi_2^\dagger \Phi_1 \right)\,\,\left(\Phi_3-\Phi_3^\star\right)  \right]
  \label{scalarpot}
  \eea
  where 
  \bea
  &&\Phi_1\equiv\left[\begin{array}{c} \phi_1^+\\
  \frac{1}{\sqrt{2}} \,\left(v_{\rm SM}+\phi_1^0 + i\,\eta_1^0\right)\\\end{array}\right];\,\, \, \Phi_2\equiv\left[\begin{array}{c} \phi_2^+\\
  \frac{1}{\sqrt{2}} \,\left( \phi_2^0 + i\,\eta_2^0\right)\\ \end{array}\right]\,\, \, {\rm and} \,\Phi_3\equiv \frac{1}{\sqrt{2}}\, \left[v_s +\phi_3^0 + i\,\eta_3^0 \right]\nn\\ \label{scalarfields}
\eea
\bea
  -{\cal L}_{\rm Y} &=& y_{u}\ \overline{Q_L}\ \widetilde{\Phi_1}\ u_R\ +\ y_{d}\ \overline{Q_L}\ \Phi_1\ d_R\ +\ y_{{}_{l}}\ \overline{l_L}\ \Phi_1 \ e_R\ + y_1\ \overline{l_L}\ \Phi_2 \ e_R  +\ \text{h.c.} \label{SMYukawa}\\
  {\cal L}_{\rm VL}&=&  \overline{\chi}\ i \left(\not\!\! \partial - i g^\prime\frac{Y}{2} \not\!\!B \right) \chi\ -\ m_{\chi}\ \overline{\chi}\ \chi -  y_2\ \overline{\chi_L}\ \chi_R\ \Phi_3 -\ y_3\ \overline{\chi_L}\ e_R\ \Phi_3 \label{VLYukawa}
 \eea
 \end{subequations}
  where all couplings in the scalar potential and Yukawa sector    are real in order to preserve the CP invariance. The quartic scalar couplings are taken to be perturbative {\i.e.} $\left\vert\lambda_i\right\vert \le 4 \pi$. Here, we have invoked an additional global $U(1)$ symmetry such that $\Phi_3 \to e^{ i\,\alpha} \Phi_3$  to reduce the number of free parameters in the scalar potential, which is however allowed to be softly broken by the $\kappa$ term  and Yukawa couplings $y_2$ and $y_3$.
  
 \subsection{Positivity and minimisation Conditions}
 In order to have a stable minimum ({\it i.e.} potential bounded from below), the parameters of the potential need to satisfy positivity
conditions which are essentially governed   by the quartic terms in the scalar potential.
  The co-positivity conditions for the Lagrangian given in \eqref{scalarpot} are obtained by demanding the determinant and principal minors of the Hessian to be positive definte. The couplings are then required to satisfy 
  \bea
  {\cal H} =\left\vert\begin{array}{ccc} \l_1 &\  \l_3+\l_4-|\l_5|\ &\  \l_{11}\\
 \l_3+\l_4-|\l_5|\ &\ \l_2 & \l_{13} \\
  \l_{11}\ &\ \l_{13}\ &\ \l_8 \end{array}\right\vert > 0\nn\\
  \eea
 along with $\l_1, \l_2\, {\rm and}\, \l_8 > 0$.
 This leads to the following co-positivity conditions:
 \begin{subequations}
\begin{eqnarray}
 &&\lambda_1, \lambda_2, \lambda_{8} > 0,\\
  &&{\bar{\lambda}_{12}} \equiv \lambda_3 + \Theta\left[\left\vert \lambda_5\right\vert -\lambda_4 \right]\ 
 (\lambda_4-|\lambda_5|) + \sqrt{\lambda_1 \lambda_2}  > 0,\label{secondcopositivity}\\
 &&{\bar{\lambda}_{13}} \equiv \lambda_{11}  + \sqrt{ \lambda_1 \lambda_{8}} > 0,\\
 &&{\bar{\lambda}_{23}} \equiv \lambda_{13}  + \sqrt{ \lambda_2 \lambda_{8}} > 0\ \text{and}\\
 && \sqrt{\lambda_1 \lambda_2\lambda_{8}} + 
[\lambda_3 + \Theta[\left\vert\lambda_5\right\vert-\lambda_4] (\lambda_4-\left\vert\lambda_5\right\vert)] \sqrt{\lambda_{8}}+
\lambda_{11} \sqrt{ \lambda_2} 
+ \sqrt{2\ {\bar{\lambda}_{12}} {\bar{\lambda}_{13}} {\bar{\lambda}_{23}} } > 0\nn\\
\end{eqnarray}
 \end{subequations}
\par Considering the VEV's  for $\Phi_1$ and $\Phi_3$ to be real, we   minimise  the scalar potential \eqref{scalarpot} which leads to the following two minimisation conditions:
\begin{subequations}
\bea
m_{11}^2&=&\l_1\ v_{\rm SM}^2\ +\ \l_{11}\ v_{s}^2\\
 m_{33}^2&=& \l_8\ v_{s}^2\ +\ \l_{11}\ v_{\rm SM}^2
 \eea
 \label{minimise}
 \end{subequations}
 The $m_{22}^2$ parameter remains unconstrained by the extremum condition. 
 \par

 %%%%%%%%%%%%%%%%%%%%%%%%%%%%%%%%%%%%
 \subsection{Scalar and Pseudo-scalar Mass eigenstates}
 The squared-mass matrix constructed from all six scalar components of the scalar fields  is given by 
 \begin{eqnarray}
M_{\phi_i\,\phi_j}^2 = \left.\frac{\partial^2V}{\partial\phi_i\,\partial \phi_j}\right\vert_{\Phi_i=\left\langle\Phi_i\right\rangle},\,\,\,{\rm for} i,\, j \equiv 1, \, \,...\, 6
 \end{eqnarray}
 with $\phi_i$ being the respective scalar and/ or pseudo-scalar  fields as defined in equation \eqref{scalarfields}. 
 \par As there is no mixing among the imaginary  component of the inert doublet with the real component of either the first SM like doublet or the singlet, the two mass matrices  for neutral scalars and pseudo-scalars are therefore completely decoupled. 
 \par The $2\times 2$  CP-even  neutral scalar  mass matrix arises due to the mixing of the real components of SM like first doublet $\Phi_1$ and the singlet $\Phi_3$ and is given as
   
 \bea
 M_{\phi_1^0\,\phi_3^0}^2=\frac{1}{2}\begin{pmatrix} \phi_1^0 & \phi_3^0 \end{pmatrix} \begin{pmatrix}  \l_1 \,\, v_{\rm SM}^2  &\ \l_{11}\,\, v_{\rm SM} \,\, v_{s} \\
 \l_{11} \,\, v_{\rm SM} \,\, v_{s}\ &\ \l_8 \,\, v_{s}^2 \end{pmatrix} \begin{pmatrix} \phi_1^0 \\ \phi_3^0 \end{pmatrix}
 \eea 
 
 On diagonalising the CP-even mass matrix by orthogonal rotation matrix parameterised in terms of the mixing angle $\theta_{13}$
 we get the two  mass eigenstates $h_1$ and $h_3$. The  mass eigenvalues  are 
 \begin{subequations}
 \bea
m_{h_1}^2&=& \cos^2 {\theta_{13}}  \,\,\lambda _1 \,\,v_{\rm SM}^2+\sin  \left(2 {\theta_{13}}\right)  v_{s}\,\, \lambda _{11} \,\,v_{\rm SM}+\sin^2 {\theta_{13}} \,\, v_{s}^2\,\, \lambda _8\\
m_{h_3}^2&=& \sin ^2 {\theta_{13}}  \,\, \lambda _1\,\, v_{\rm SM}^2-\sin  \left(2 {\theta_{13}} \right) v_{s} \,\,\lambda _{11} \,\, v_{\rm SM}+\cos^2 {\theta_{13}} \,\, v_{s}^2\,\, \lambda _8
\eea 
 The vanishing off diagonal term of the diagonalised mass matrix defines the mixing angle in terms of other model parameters as follows:
\bea
 \tan 2\theta_{13}&=&\frac{\lambda_{11}\ v_{\rm SM} v_{s}}{\l_1 v_{\rm SM}^2-\l_8 v_{s}^2}
 \eea
 \end{subequations}
\par Similarly, we diagonalise the following mass matrix for CP-odd scalars $\eta_2^0$ and  $\eta_3^0$ by the orthogonal rotation matrix parameterised by mixing angle $\theta_{23}$
\bea
 \frac{1}{2}\begin{pmatrix} \eta_2^0\ &\ \eta_3^0 \end{pmatrix} \begin{pmatrix} -\frac{1}{2} m_{22}^2+ \frac{1}{2}\overline{\l}_{345} v_{\rm SM}^2 + \frac{1}{2}v_{s}^2 \l_{13}\ &\ -\sqrt{2}  \kappa  v_{\rm SM}\\
 -\sqrt{2}  \kappa  v_{\rm SM}\ &\ 0 \end{pmatrix} \begin{pmatrix} \eta_2^0 \\ \eta_3^0 \end{pmatrix}
 \eea
where $\overline{\lambda}_{345}=\lambda_3+\lambda_4-\lambda_5$.  
The mass eigenvalues of the pseudo-scalar mass eigenstates $A^0$ and $P^0$ are calculated to be 
\begin{subequations}
\bea
m_{A^0}^2&=& \frac{1}{2} \left( \overline{\lambda}_{345} v_{\rm SM}^2- m_{22}^2 +\l_{13} v_{s}^2\right)\cos^2\theta_{23}- \sqrt{2} \kappa v_{\rm SM} \sin2\theta_{23}\\
m_{P^0}^2&=& \frac{1}{2} \left( \overline{\lambda}_{345} v_{\rm SM}^2- m_{22}^2 +\l_{13} v_{s}^2\right)\sin^2\theta_{23}+ \sqrt{2} \kappa v_{\rm SM} \sin2\theta_{23}
\eea 
The off-diagonal vanishing  terms relates the mixing angle $\theta_{23}$ to the other mass and model parameters as:
\begin{eqnarray}
\kappa = -\,\frac{1}{2\,\sqrt{2} v_{\rm SM}}\left(m^2_{P^0}+m^2_{A^0}\right)\tan\left(2\,\theta_{23}\right)
\end{eqnarray}
\end{subequations}
 \par Defining the remaining neutral and charged saclar mass eigenstates as
 \bea
 \phi_2^0\ &\to& h_2\nn\\
 \eta_1^0\ &\to&\ G^0\  (\text{massless Nambu-Goldstone Boson})\nn\\
 \phi_1^\pm\ &\to&\ G^\pm\  (\text{massless Nambu-Goldstone Boson})\nn\\
 \phi_2^\pm\ &\to&\ H^\pm\nn
 \eea
with
\begin{subequations}
 \bea
 m_{h_2}^2&=&\frac{1}{2}\ \left[  -m_{22}^2+\  \left(\lambda _3 + \lambda _4 + \lambda _5\right)\ v_{\rm SM}^2 + \l_{13} v_{s}^2\right]\\
 m_{H^\pm}^2&=&- m_{22}^2\ + \lambda_3\ v_{\rm SM}^2+\ \l_{13} v_{s}^2
 \eea
 \end{subequations}
 \par The twelve independent parameters in the scalar potential \eqref{scalarpot}, 
\bea
m_{11},\ m_{22},\ m_{33},\ \lambda_{i= 1,\,\ldots \, 5},\ \lambda_8,\ \lambda_{11},\ \lambda_{13}\ \text{and}\ \kappa
\eea
can now be expressed in terms of the following physical masses and mixing angles:
  \bea
  v_{\rm SM},\, v_s,\, m_{h_1}^2,\, m_{h_1}^2,\, m_{h_2}^2,\, m_{H^\pm}^2,\,\ m_{A^0}^2,\, m_{P^0}^2,\, \theta_{13},\, \theta_{23}\, \text{and}\ m_{22}^2
  \eea
 These mass relations are given in the Appendix \ref{App:InputParams}. 
 
\par Substituting the mass relations of $\l_4$ and $\l_5$ from equations \eqref{lam4} and \eqref{lam5} respectively in the theta function appearing in equation \eqref{secondcopositivity} of the co-positivity conditions.
  we get two mutually exclusive allowed regions of parameter space:
  \bea
  \Theta(\left\vert \lambda_5\right\vert-\lambda_4)=\left\{\begin{array}{cc} \Theta\left[m_{H^\pm}^2-(m_{A^0}^2+m_{P^0}^2)\right] &\,\,\text{for}\,\,\,\, m_{h_2}^2>m_{A^0}^2+m_{P^0}^2 \\
   \Theta\left[m_{h_2}^2-m_{H^\pm}^2\right]
     & \,\,\text{for}\,\,\,\, m_{h_2}^2<m_{A^0}^2+m_{P^0}^2 \\
  \end{array}\right.\nn\\
 \label{derivedpositivity}
  \eea
In this article we explore the  phenomenologically interesting  region where $m_{H^\pm}^2>m_{A^0}^2+m_{P^0}^2$.
\subsection{Yukawa Couplings} 
 The Yukawa interactions given in \eqref{SMYukawa} and \eqref{VLYukawa}  can be re-written as 
 \begin{subequations}
  \bea
 -{\cal L}^{\rm \small Yukawa}_{\rm \small SM\, Fermions}&=&\sum_{s_i\equiv h_1,h_3}\frac{y_{ffs_i}}{\sqrt{2}} \left(v_{\rm SM}\,\,\delta_{s_i,h_1} + s_i\right) \bar f\ f\ + \frac{y_{llh_2}}{\sqrt{2}} h_2\ \bar l^- \ l^- \sum_{s_i\equiv P^0,A^0} \frac{y_{lls_i}}{\sqrt{2}} s_i\ \bar l^-  \gamma_5\ l^-\nn\\
 && +\ \left[y_{l\nu H^-}\  \bar \nu_l\ P_R\ l^-   H^+ +  {\rm h.c.} \right]\\
-{\cal L}^{\rm Yukawa}_{\rm VL\ Leptons}&=&
\sum_{s_i\equiv h_1,h_3,A^0,P^0}\frac{1}{\sqrt{2}} \left(v_s\,\, \delta_{s_i,h_3}+ s_i\right)\ \bar \chi \left(y_{\chi\chi s_i} P_R+  y_{\chi\chi s_i}^\star P_L\right) \chi\\
 -{\cal L}^{\rm Yukawa}_{\rm VL,\ SM\  Leptons}&=& \sum_{s_i\equiv h_1,h_3,A^0,P^0}\frac{1}{\sqrt{2}} \left(v_s\,\, \delta_{s_i,h_3}+ s_i\right) \left[y_{l\chi s_i}\ \bar \chi\  P_R\, l^- + {\rm h.c.}\right]
 \eea
 \label{yukawamasseigenstates}
\end{subequations}
where $f$ and  $l^-$ represent SM fermions and SM charged leptons respectively. The Yukawa couplings $y_{\psi_1\psi_2S_i}$ with scalar/ pseudoscalar mass eigenstates  are given in Table \ref{Table:Yukawa}.  
 \begin{table}[htb]\footnotesize
 \begin{center}
 \begin{tabular}{|c|c||c|c|}
 \hline
 \hline
$y_{ffh_1}$ & $\left(\sqrt{2} m_f/ v_{\rm SM}\right) \cos\theta_{13}$   &  $y_{llh_2}$&$y_1$\\
 \hline
$y_{ffh_3}$&$- \left(\sqrt{2}m_f/ v_{\rm SM}\right)  \sin\theta_{13}$  & $y_{llP^0}$&$-i\ y_1 \sin\theta_{23}$\\
 \hline
 $y_{\chi\chi h_1}$&$y_2\   \sin\theta_{13}$  &$y_{llA^0}$&$i\ y_1 \cos\theta_{23}$
\\
\hline
$y_{\chi\chi h_3}$&$y_2\  \cos\theta_{13}$ &$y_{l\chi h_1}$&$y_3 \sin\theta_{13}$\\
\hline
$y_{\chi\chi P^0}$&$i\ y_2\  \cos\theta_{23}$ &$y_{l\chi h_3}$&$y_3 \cos\theta_{13}$\\
\hline
 $y_{\chi\chi A^0}$&$i\ y_2\ \, \sin\theta_{23}$ &$y_{l\chi P^0}$&$i\ y_3 \cos\theta_{23}$\\
 \hline
 $y_{l\nu H^-}$ &$y_1$ &$y_{l\chi A^0}$ &$i\ y_3 \sin\theta_{23}$\\\hline
 \end{tabular}
 \end{center}
 \caption{\em{Yukawa couplings}}
 \label{Table:Yukawa}
\end{table}

%%%%%%%%%%%%%%%%%%%%%%%%%%%%%%%%%%%%%%%%%%%%%%%%%%%%%%%%%%%%%%%%%%%%%%%%%%%%%%%%%%%%%%%%%%%%%%%%%%%%%%
%%%%%%%%%%%%%%%%%%
\section{Experimental Constraints}
\label{sec:EWconstraints}
Any model beyond the SM has to satisfy the existing theoretical and experimental observations. In this section, we subject the model discussed in section \ref{sec:model} to the observations of SM-like Higgs mass and signal strengths as measured at the LHC Run-II and at the ILC. We further examine the electroweak precision constraints on the masses of scalars and pseudo-scalars  from the direct production at LEP-II.

\subsection{Higgs decays to Gauge Bosons}
Any multi-Higgs model has to accommodate the SM like Higgs with the mass and signal strengths measured at the LHC \cite{PDG:2020} with the future prospects of increasing precision measurements at the future collider experiments. We identify and align the  CP even lightest scalar mass eigenstate $h_1$ with $125.09$ GeV SM Higgs. Therefore, the  couplings of the  $h_1$ with a pair of fermions and gauge bosons are essentially those of SM Higgs couplings but suppressed by  $\cos\theta_{13}$ due to $\Phi_1-\Phi_3$ small angle mixing ($\theta_{13} = 0$ restores the full SM Higgs).
\begin{figure}[h!]
\centering
\includegraphics[width=0.55\textwidth,clip]{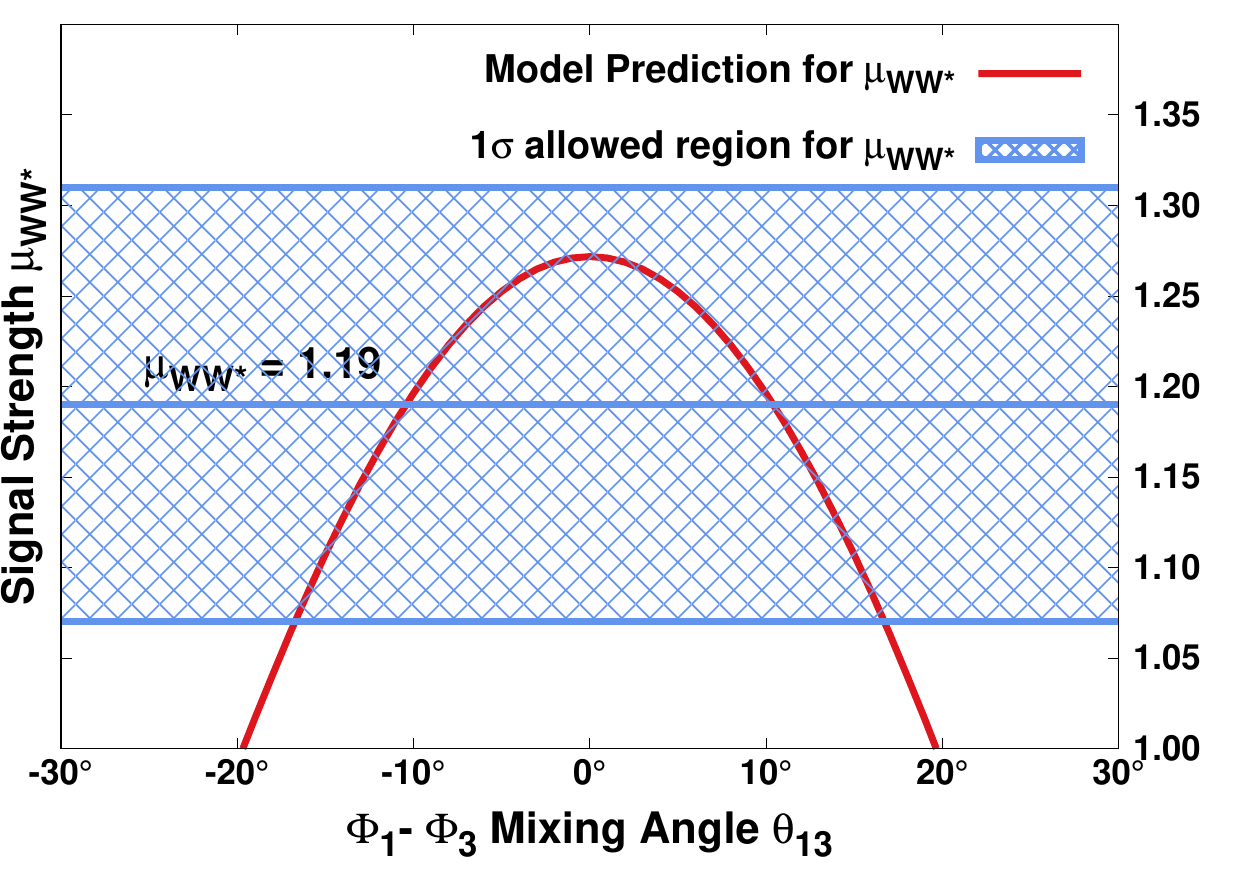}
\caption{ \em{ Variation of  $\mu_{{}_{W W^\star}}$ with the CP-even mixing angle $\theta_{13}$ is shown with the dashed red line. The allowed shaded one sigma region for measured $\mu_{{}_{W W^\star}} = 1.19 \pm 0.12$ \cite{PDG:2020} is also shown.}}
\label{fig:Hdecay1}
\end{figure}

\par We compare the total Higgs decay width in SM  $\Gamma(h^{\rm SM}\to \textrm{all})\sim 4.07$ MeV \cite{Denner:2011mq, LHCHiggsCrossSectionWorkingGroup:2013rie} with the recently measured total decay width    $\Gamma(h_1\to \textrm{all}) \sim 3.2^{+2.8}_{-2.2}$ MeV at the LHC \cite{PDG:2020}. Assuming, that  the  model can account for the measured value of the total decay width, we constrain the model parameters by examining the bounds on the partial decay widths of decay channels for 125 GeV $h_1$ at LHC. To this end,  we define the signal strength  $\mu_{{}_{XY}}$ {\it w.r.t.} $h_1$ production {\it via} dominant gluon fusion in $p-p$ collision, followed by its  decay to  $X\ \&\ Y$ pairs in the narrow width approximation as
\bea
\mu_{{}_{XY}}&=&\frac{\sigma(pp\to h_1\to XY)}{\sigma(pp\to h\to XY)^{\textrm  {SM}}}=\frac{\Gamma\left( h_1\to g\, g\right)}{\Gamma \left(h^{SM} \to g\, g\right)}\,\,\,\, \frac{\textrm{BR}\left(h_1\to X\,Y\right)}{\textrm{BR}\left(h^{\rm SM}\to X\,Y\right)}\nn\\
&=&\cos^4\theta_{13}\,\,\frac{\Gamma(h^{\rm SM}\to \textrm{all})}{\Gamma(h_1\to \textrm{all})}
\eea
\par We first analyse the partial decay width of $h_1\to W\,W^\star$ channel which solely depends   on $\theta_{13}$:
\bea
\Gamma(h_1\to W W^\star)&=& \cos^2 \theta_{13}\ \Gamma(h^{\rm SM} \to W W^\star)
\label{h1toww}
\eea
 Among the signal strengths for Higgs decaying to gauge Bosons at tree level,  $\mu_{W W^\star}\sim 1.19 \pm 0.12$ \cite{PDG:2020} has the least uncertainty for which it can provide the strongest upper bound on the mixing angle $\theta_{13}$.  In figure \ref{fig:Hdecay1}, we show the one sigma band around the central value of the $\mu_{W W^\star} $ which restricts  the model contribution curve drawn  in green for  $\left\vert\theta_{13}\right\vert \lesssim 10^\circ$ at 1 $\sigma$ level.
 
\par Next, we calculate the partial  decay widths generated at one loop for $h_1\to\gamma\,\gamma$ and $h_1\to Z\,\gamma$ respectively. The contribution of charged scalars $H^\pm$ and vector-like leptons $\chi^-$ modify the SM predictions for the branching ratios. The partial decay widths for $\gamma\,\gamma$ and $Z\,\gamma$ in the model are parameterised as
\begin{subequations}
\begin{eqnarray}
\Gamma(h_1\to\gamma\,\gamma)&=& \cos^2\theta_{13} \left\vert 1 + \zeta_{\gamma\gamma}\right\vert^2\,\Gamma\left( h^{\rm SM} \to \gamma\,\gamma\right)
\label{h1toaa}\\
\Gamma(h_1\to Z\,\gamma)&=& \cos^2\theta_{13} \left\vert 1 + \zeta_{Z\gamma}\right\vert^2\,\Gamma\left( h^{\rm SM} \to Z\,\gamma\right)\label{h1toza}
\end{eqnarray}
\end{subequations}
where the SM Higgs partial decay widths in $\gamma\,\gamma$ and $Z\,\gamma$ channels are given as
\begin{subequations}
\bea
\Gamma(h^{\rm SM}\to\gamma\gamma)&=&\frac{G_F\alpha^2\ m_{h}^3}{128\sqrt{2}\pi^3}\ \left\vert  \frac{4}{3}  {\cal  M}^{\gamma\gamma}_{1/2}\bigg(\frac{4m_t^2}{m_{h}^2}\bigg)+  {\cal  M}^{\gamma\gamma}_1\bigg(\frac{4m_{\rm W}^2}{m_{h}^2}\bigg) \right\vert^2\label{hsmtoaa}\\
\Gamma (h^{\rm SM }\ra Z\gamma ) &=& \frac{G^2_{F}\ \alpha\,m_W^2\, m_{h}^{3}} 
{64\,\pi^{4}} \left( 1-\frac{m_Z^2}{m_h^2} \right)^3 \left\vert
2 \frac{\left( 1- \frac{8}{3} s^2_W \right)}{c_W} {\cal  M}^{Z\gamma}_{1/2}\left(\frac{4 m_t^2}{m_h^2},\frac{4 m_t^2}{m_Z^2}\right) + 
{\cal  M}^{Z\gamma}_1\left(\frac{4 m_W^2}{m_h^2},\frac{4 m_W^2}{m_Z^2}\right) \right\vert^2 \nn\\
\label{hsmtoza}
\eea
\end{subequations}
\begin{figure}[h!]
\centering
\begin{multicols}{2}
\includegraphics[width=0.5\textwidth,clip]{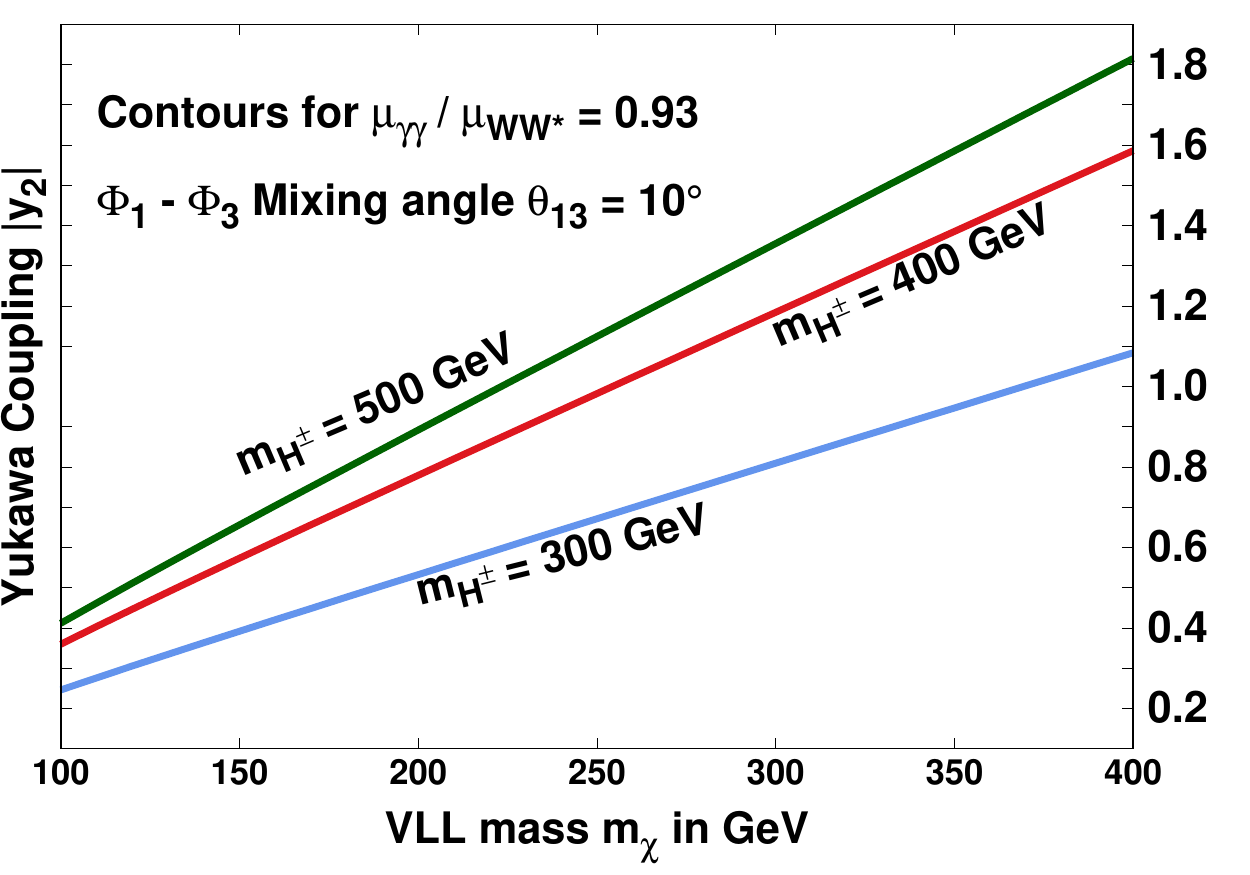}
\subcaption{\small \em{  }}\label{fig:Hdecay2a}
\includegraphics[width=0.5\textwidth,clip]{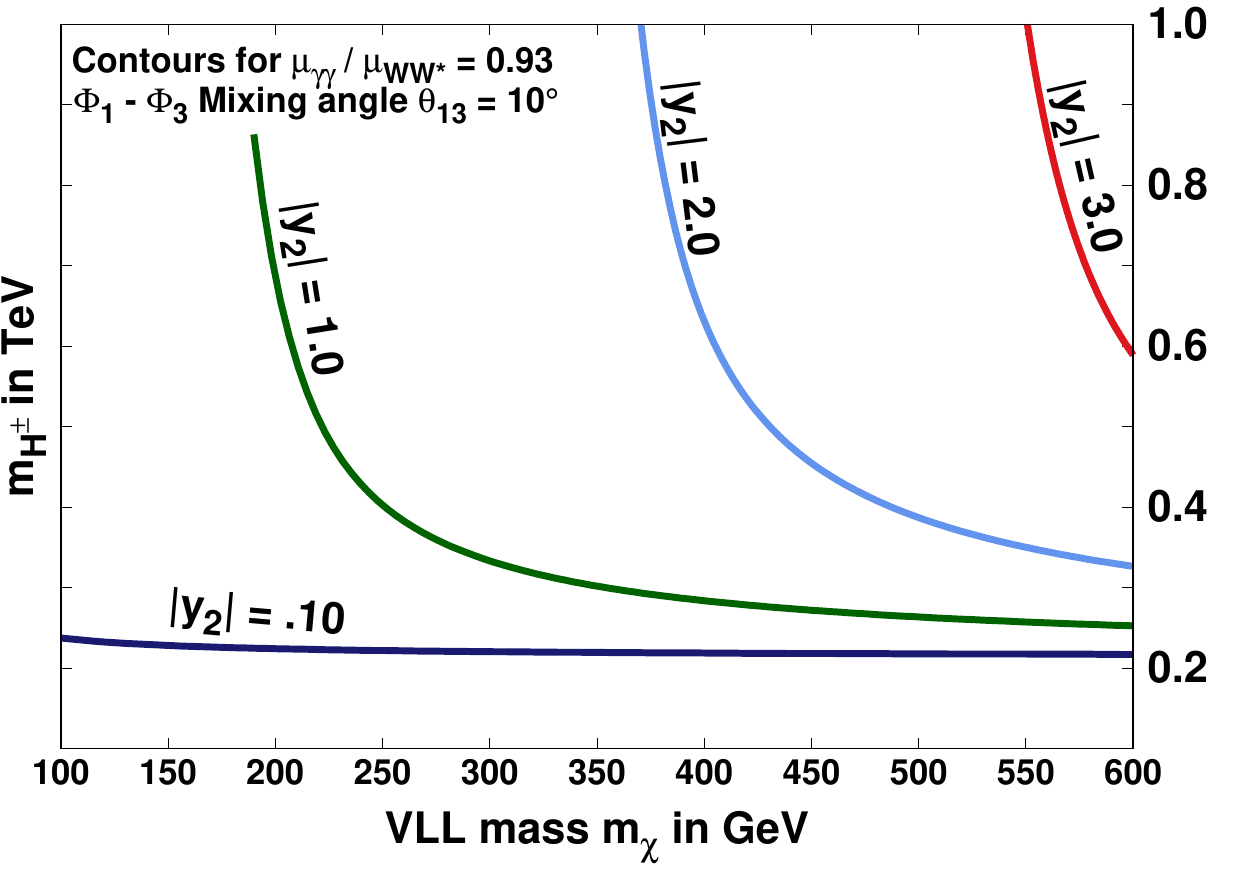}
\subcaption{\small \em{} }\label{fig:Hdecay2b}
\end{multicols}
\caption{ \em{ Contours satisfying the constraint $\mu_{{}_{\gamma \gamma}}/\mu_{{}_{W W^\star}}= 0.93^{ +0.17}_{ -0.16}$ \cite{PDG:2020} for $\theta_{13} = 10^\circ$ are drawn in   (a)  $m_\chi-\left\vert y_2\right\vert$ plane corresponding to three choices of charged Higgs mass and (b) $m_\chi - m_{H^\pm}$ plane corresponding to four choices of $\left\vert y_2\right\vert$. }}
\label{fig:Hdecay2}
\end{figure}
The dimensionless parameters $\zeta_{\gamma\gamma}$ and $\zeta_{Z\gamma}$ are defined as 
\begin{subequations}
\bea
\zeta_{\gamma\gamma} &=& \frac{v_{\rm SM}}{\cos\theta_{13}}\left[\frac{\frac{g_{h_1H^+H^-}}{2\,m^2_{H^\pm}} {\cal M}^{\gamma\gamma}_0\left(\frac{4 m_{H^\pm}^2}{ m_{h_1}^2} \right)  + \frac{y_2}{\sqrt{2} m_\chi} \sin\theta_{13} {\cal M}^{\gamma\gamma}_{1/2}\left(\frac{4\,m_{\chi}^2}{ m_{h_1}^2}\right)}{{\cal M}^{\gamma\gamma}_1\left(\frac{4\,m_{\rm W}^2} {m_{h_1}^2}\right) + \frac{4}{3}\, {\cal M}^{\gamma\gamma}_{1/2}\left(\frac{4\,m_t^2}{m_{h_1}^2}\right)}\right]\label{zetaaa}\\
\zeta_{Z\gamma} &=& \frac{v_{\rm SM}}{\cos\theta_{13}}\left[\frac{-\,\frac{ g_{h_1 H^+ H^-}}{2 m_{H^\pm}^2}\, \frac{1- 2 s^2_{\theta_W}}{c_{\theta_W}} {\cal M}^{Z\gamma}_0\left(\frac{4 m_{H^\pm}^2}{m_{h_1}^2},\frac{4 m_{H^\pm}^2}{m_Z^2}\right)\ -\  \frac{y_2}{\sqrt{2} m_\chi}\ \sin\theta_{13}\ \frac{4\ s^2_W}{c_W} \,{\cal M}^{Z\gamma}_{1/2}\left(\frac{4 m_\chi^2}{m_{h_1}^2},\frac{4 m_\chi^2}{m_Z^2}\right)}
{2 \frac{\left( 1- \frac{8}{3} s^2_W \right)}{c_W} {\cal M}^{Z\gamma}_{1/2}\left(\frac{4 m_t^2}{m_{h_1}^2},\frac{4 m_t^2}{m_Z^2}\right) + {\cal M}^{Z\gamma}_{1}\left(\frac{4 m_W^2}{m_{h_1}^2},\frac{4 m_W^2}{m_Z^2}\right)}\right]\nn\\
\label{zetaza}
\eea
\end{subequations}
where the triple scalar coupling  $ g_{h_1 H^+ H^-}  = \left(v_{\rm SM} \,\lambda_3 \,\cos\theta_{13} +  v_s\,\lambda_{13}\,\sin\theta_{13}\right)$.     The form factors ${\cal M}^{\gamma\gamma}_{0,\,1/2,\,1}$ and ${\cal M}^{Z\gamma}_{0,\,1/2,\,1}$ are defined in the appendix \ref{app:formfactors}.   The analytical expressions  for the partial widths except the additional contribution from VLL are identical to that given in the reference \cite{Bonilla:2014xba}. 
\par Using the analytical expressions for the partial decay widths   for $h_1\to \gamma\, \gamma$, $h_1\to Z\, \gamma$ and $h_1\to W \,W^\star$   in equations \eqref{h1toaa}, \eqref{h1toza} and \eqref{h1toww} respectively, we consider the two ratios of the signal strengths namely, 
\begin{subequations}
\bea
\frac{\mu_{{}_{\gamma \gamma}}}{\mu_{{}_{W \,W^\star}}}&=& \frac{\Gamma(h_1 \to \gamma \,\gamma)}{\Gamma(h_1 \to W\, W^\star)} \times \frac{\Gamma(h^{\rm SM} \to W W^\star)}{\Gamma(h^{\rm SM} \to \gamma \,\gamma)}=\left\vert 1 + \zeta_{\gamma\gamma} \right\vert^2
\label{aabywratio}\\
\frac{\mu_{{}_{Z \,\gamma}}}{\mu_{{}_{W\, W^\star}}}&=& \frac{\Gamma(h_1 \to Z \,\gamma)}{\Gamma(h_1 \to W \,W^\star)} \times \frac{\Gamma(h^{\rm SM} \to W\, W^\star)}{\Gamma(h^{\rm SM} \to Z\, \gamma)}= \left\vert 1 + \zeta_{Z\gamma} \right\vert^2 \label{zabywratio}
\eea
\end{subequations}
The experimental limits on   $\mu_{{}_{\gamma \gamma}}=1.11^{+0.10}_{-0.09}$ and $\mu_{{}_{W W^\star}}=1.19 \pm 0.12$ for $m_{h^{\rm SM}}=125.09$ GeV \cite{PDG:2020} are then substituted in equations \eqref{aabywratio} to constrain the allowed parameter space for  the Charged Higgs and singlet vector like lepton  respectively. Since experimental uncertainty for $\mu_{{}_{Z \gamma }}<6.6$ \cite{PDG:2020} is large, we do not constrain the model parameters from $h_1\to Z\,\gamma$ decay channel.

 \par We depict the contours satisfying $\mu_{\gamma \gamma}/ \mu_{W W^\star} =0.93^{+0.17}_{-0.16}$ for  $\theta_{13} = 10^\circ$  in  figure \ref{fig:Hdecay2}. In figure \ref{fig:Hdecay2a} contours satisfying the central value of the ratio are drawn in the $m_\chi - \left\vert y_2\right\vert$  plane for $m_{H^\pm}$ = 300, 400 and 500 GeV.  The sensitivity of the charged Higgs mass are studied {\it w.r.t.} variation of the VLL mass from the contours satisfying the central value of the ratio in figure  \ref{fig:Hdecay2b} for four choices of Yukawa couplings $\left\vert y_2\right\vert$ varying between 0.1 and the perturbative limit $\sqrt{4\pi}$.

%%%%%%%
\begin{figure}[h!]
\centering
\begin{multicols}{2}
\includegraphics[width=0.5\textwidth,clip]{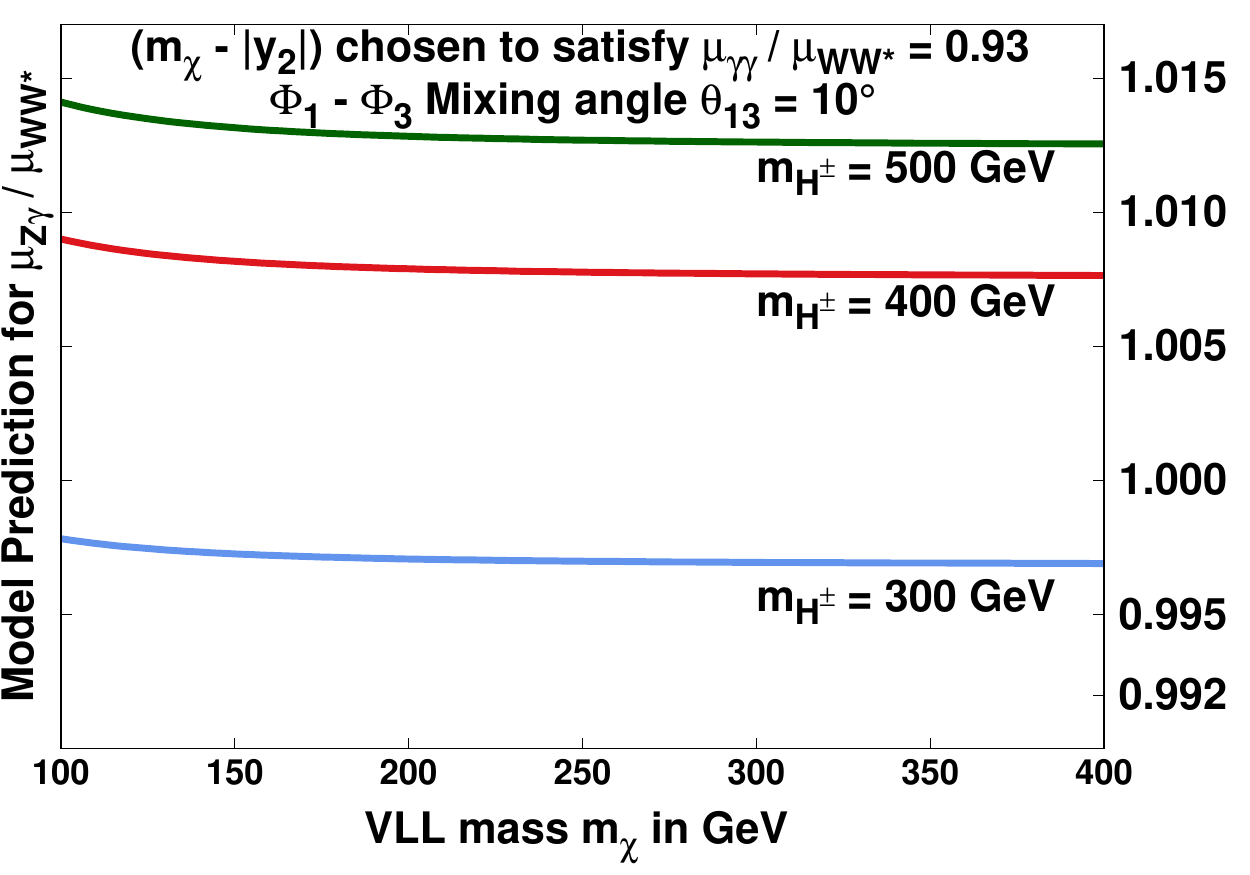}
\subcaption{ \em{  }} \label{fig:muZa}
\includegraphics[width=0.5\textwidth,clip]{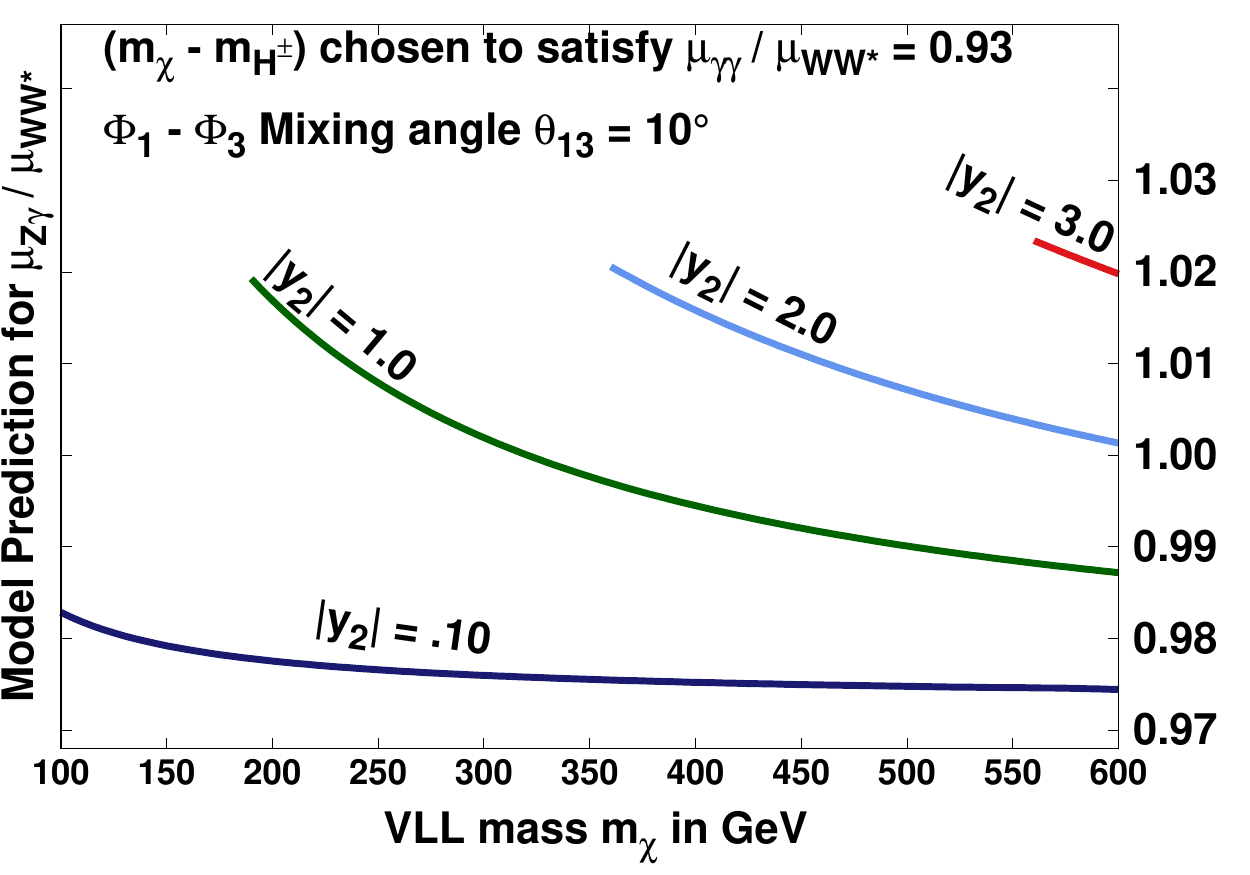}
\subcaption{ \em{} }\label{fig:muZb}
\end{multicols}
\caption{ \em{ Variation of the ratio $\mu_{Z \gamma}/\mu_{W W^\star}$   with varying $m_\chi$ for $\theta_{13}=10^\circ$. The values of $\left(m_\chi,\,\left\vert y_2\right\vert\right)$ in figure (a) and that of $\left(m_\chi,\,m_{H^\pm}\right)$ in figure (b) are such that they satisfy the constraint $\mu_{\gamma \gamma}/ \mu_{W W^\star} =0.93^{+0.17}_{-0.16}$ \cite{PDG:2020} as depicted in figures \ref{fig:Hdecay2a} and \ref{fig:Hdecay2b} respectively.
 }}
\label{fig:muZ}
\end{figure}

\par  With the constrained parameter space from $\mu_{\gamma \gamma}$ and $\mu_{W W^\star}$,  we  estimate the ratio of  signal strengths $\mu_{Z \gamma}/\mu_{W W^\star}$ and make a prediction for the future improved measurements with  higher luminosity and  kinematic reach at FCC-hh. In figure \ref{fig:muZ}  we study the variation of the ratio $\mu_{Z \gamma}/ \mu_{W W^\star}$  with varying $m_\chi$ for fixed $\theta_{13}=10^\circ$. The curves in figure \ref{fig:muZa} are depicted using  those $\left(m_\chi,\,\left\vert y_2\right\vert\right)$  values which satisfy  the constraint $\mu_{\gamma \gamma}/ \mu_{W W^\star} =0.93^{+0.17}_{-0.16}$ as shown in the 
 figure \ref{fig:Hdecay2a} corresponding to $m_{H^\pm}$ = 300, 400 and 500 GeV respectively. Similarly, the graphs in figure \ref{fig:muZb} are depicted using those $\left(m_\chi,\,m_{H^\pm}\right)$  values which satisfy the constraint $\mu_{\gamma \gamma}/ \mu_{W W^\star} =0.93^{+0.17}_{-0.16}$ as shown in the 
 figure \ref{fig:Hdecay2b} corresponding to $\left\vert y_2\right\vert  $ = 0.1, 1,0, 2.0 and  3.0 respectively.
\begin{figure}[h!]
\centering
\begin{multicols}{2}
\includegraphics[width=0.5\textwidth,clip]{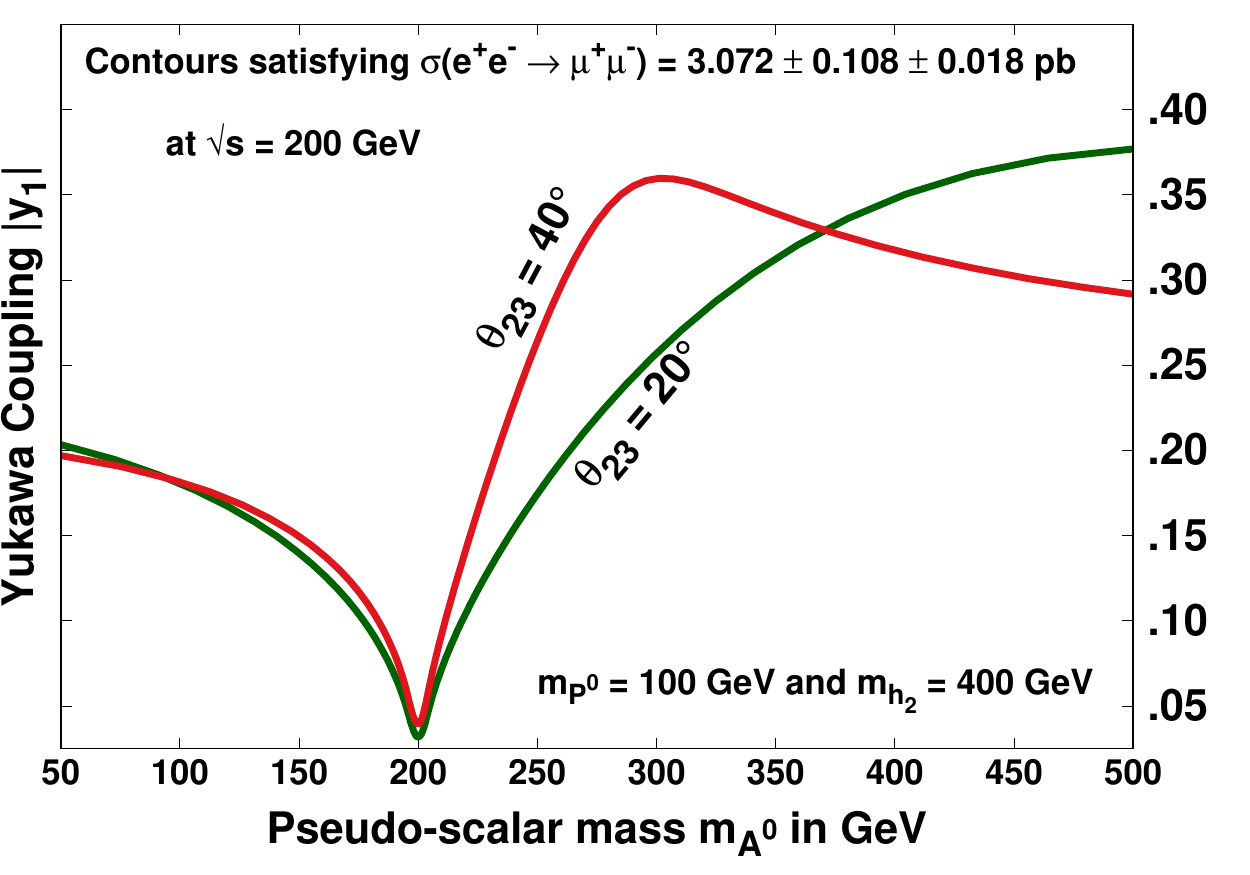}
\subcaption{\em{ $m_{P^0}=100\ GeV,\ m_{h_2}=400\ GeV.$ }}\label{fig:LEPa}
\includegraphics[width=0.5\textwidth,clip]{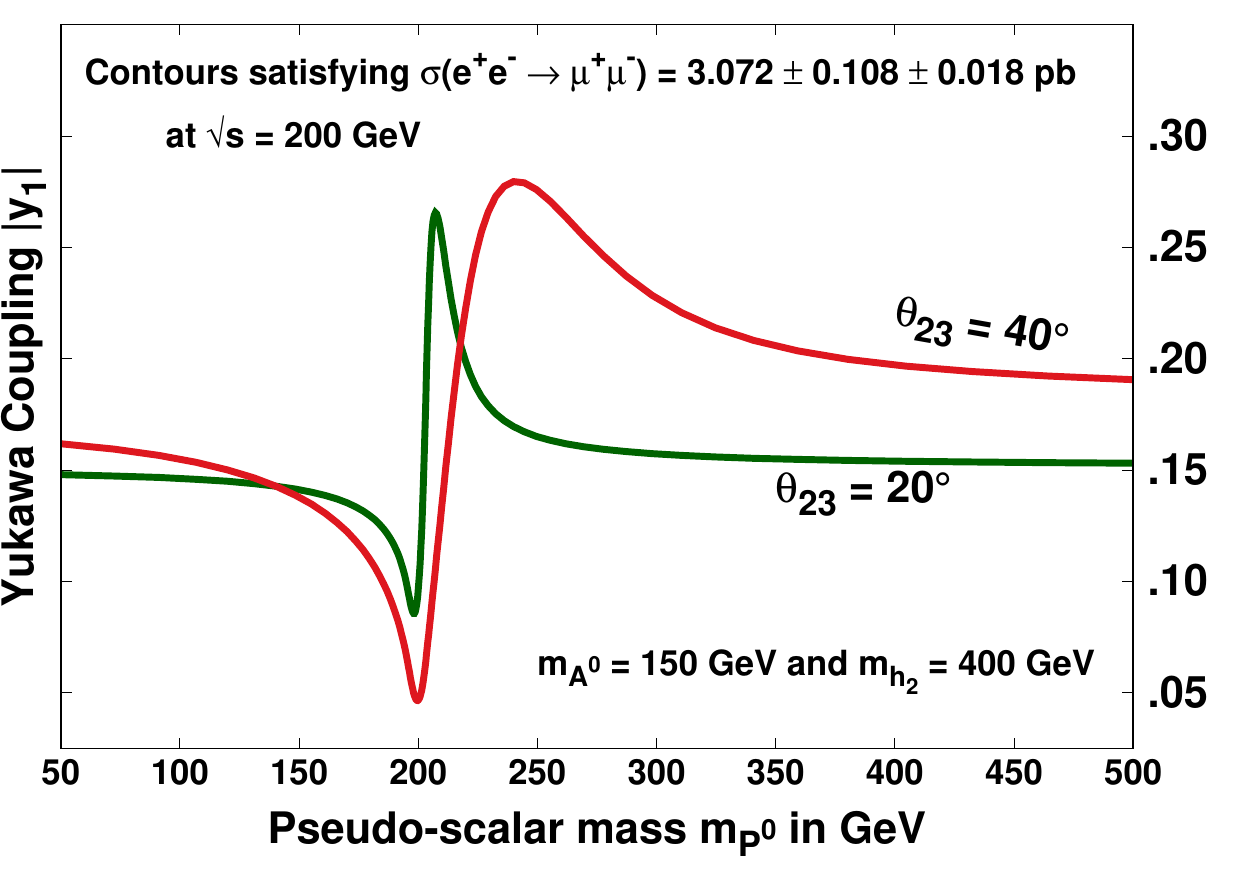}
\subcaption{\em{$m_{A^0}=150\ GeV,\ m_{h_2}=400\ GeV.$} }\label{fig:LEPb}
\end{multicols}
\begin{center}
\includegraphics[width=0.5\textwidth,clip]{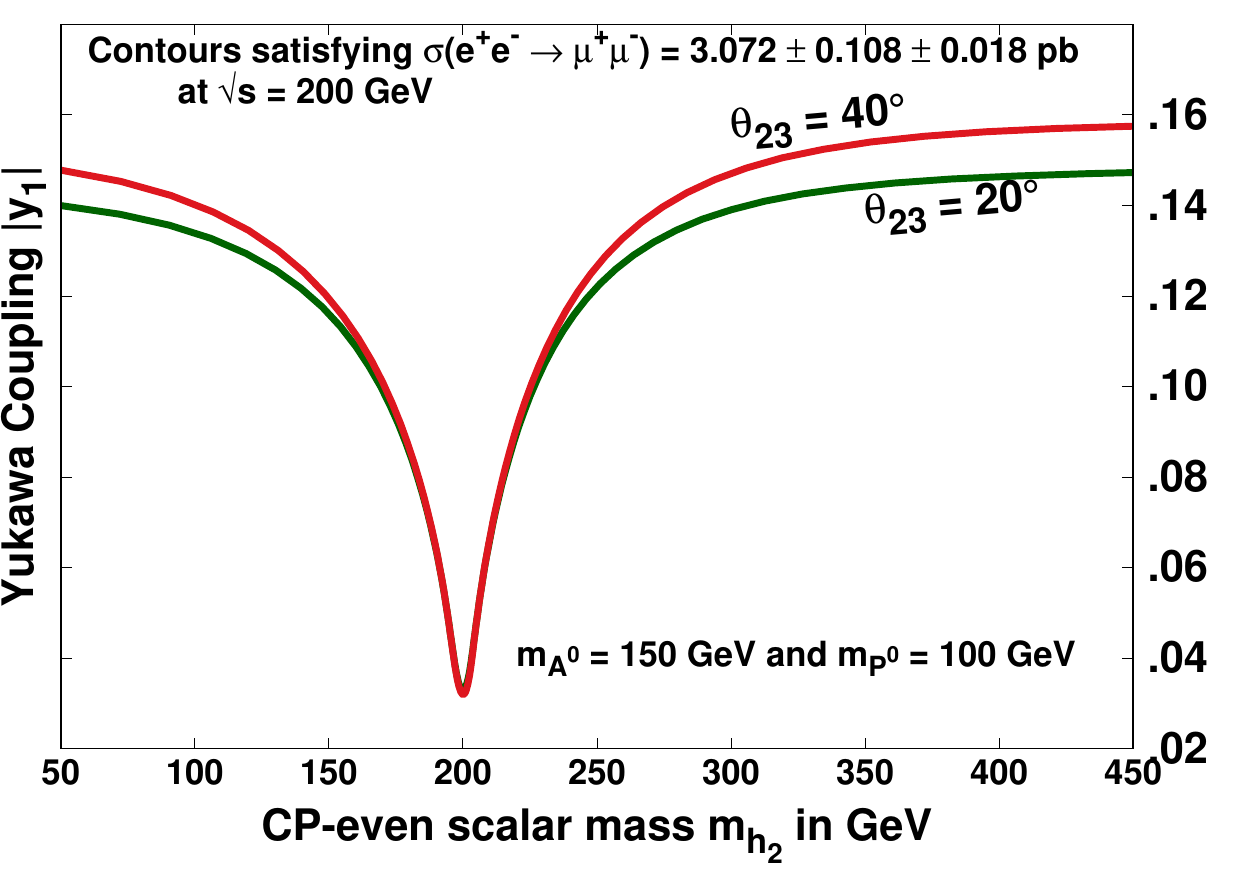}
\subcaption{\em{$m_{A^0}=150\ GeV,\ m_{P^0}=100\ GeV.$} }\label{fig:LEPc}
\end{center}
\caption{ \em{Contours in the plane defined by scalar/ pseudo-scalar mass and its coupling $\left\vert y_1\right\vert $ with SM leptons satisfying the combined analysis of DELPHI and L3 $\sigma (e^+\ e^- \to \mu^+\ \mu^-)=3.072 \pm 0.108 \pm 0.018$ pb at $\sqrt{s}=200\ GeV$  \cite{Schael:2013ita}  for three benchmark points corresponding to two choices of the mixing angle $\theta_{23}$ = 20$^\circ$ and 40$^\circ$ respectively. The parameter space lying above the contour are  forbidden by LEP observations.}}
\label{fig:LEP}
\end{figure}
%%%%%%%%%%%%%%%%%%%%%%%%
  
\subsection{LEP II Data}
We validate the model from the existing LEP II data and put the lower mass bounds on the scalar and pseudo-scalar mass eigenstates. This can be achieved either by investigating the (a) direct pair production of scalars and pseudo-scalars  or  (b) by production of pair of fermions mediated by these additional exotic physical scalars or pesudo-scalars.  VLL below 100 GeV cannot  be constrained by LEP experiment as they do not couple to  SM particles except $h_1$ (SM Like Higgs)  at tree level and therefore the cross-section is expected to be highly suppressed due to the electron Yukawa coupling.

\par The dominant direct pair production channels at $e^+e^-$ collider:
\bea
e^+\ e^- &\to& \gamma^\star/\, Z^\star \to H^+ + H^-\\
{\rm and}\hskip 0.5cm e^+\ e^- &\to& Z^\star \to A^0 / P^0 + h_i 
\eea
have been studied to put the lower mass bounds on $m_{H^\pm}\gsim$ 93.5 GeV and  $\sum m_{h_i}\gsim 200$ GeV \cite{Schael:2013ita}.

\par We compute the cross-section for $e^+\ e^- \to \mu^+\ \mu^-$ from the combined analysis at LEP II which is found to be $\sigma (e^+\ e^- \to \mu^+\ \mu^-)=3.072 \pm 0.108 \pm 0.018\ pb$ at $\sqrt{s}=200\ GeV$. 
\par The additional contribution to  $\mu$-pair production cross-section in our model can be written as:
\bea
\sigma_{\mu^+ \mu^-}^{\rm NP}&\simeq&\frac{s}{64 \pi}\ \sqrt{\frac{1-4\frac{ m_\mu^2}{s}}{1-4\frac{m_e^2}{s}}}\,y_1^4\, \left[ \left\{\frac{\cos^2\theta_{23}}{s-m_{A^0}^2} + \frac{\sin^2\theta_{23}}{s-m_{P^0}^2}\right\}^2 + \frac{1}{\left(s-m_{h_2}^2\right)^2}\right]
\label{xsec:mumu}
\eea
where we have dropped the interference terms of $h_2$  with  $h_1$ and $h_3$
as they are suppressed by  $\left(m_e\ m_\mu\right)/ v_{\rm SM}^2$. Also, the contributions from $h_1$ and $h_3$ and their interference are   suppressed by factor of $\left(m_e\ m_\mu\right)^2/v_{\rm SM}^4$ and hence have not been taken into account. The other interference terms with $\gamma^\star/\,Z$ vanish. The interference term among scalars and pseudo-scalars vanish.

 \par We compare the measured cross-section from the combined analysis of DELPHI and L3 at LEP II $\sigma (e^+\ e^- \to \mu^+\ \mu^-)=3.072 \pm 0.108 \pm 0.018$ pb at $\sqrt{s}=200\ GeV$ \cite{Schael:2013ita} with the model contribution as calculated in equation \eqref{xsec:mumu}  for three benchmark points of the parameter space. For each such bench mark point we plot  two contours  corresponding to the mixing angle $\theta_{23} = 10^\circ$  and  40$^\circ$ respectively in (a) $m_{A^0}-\left\vert y_1\right\vert$ plane, (b) $m_{P^0}-\left\vert y_1\right\vert$ plane and 
 (c) $m_{h_2}-\left\vert y_1\right\vert$ plane in figures \ref{fig:LEPa}, \ref{fig:LEPb} and \ref{fig:LEPc} respectively. The parameter space  below the respective contour is  allowed from LEP measurements.

%%%%%%%%%%%%%%%%%%%%%%%%%%%%%%%%%%%%%%%%%%%%%%%%%%%%%%%%%%%

 \subsection{Electroweak Precision Observables}
 
\begin{figure*}[h!]
\centering
\begin{multicols}{2}
\includegraphics[width=0.5\textwidth,clip]{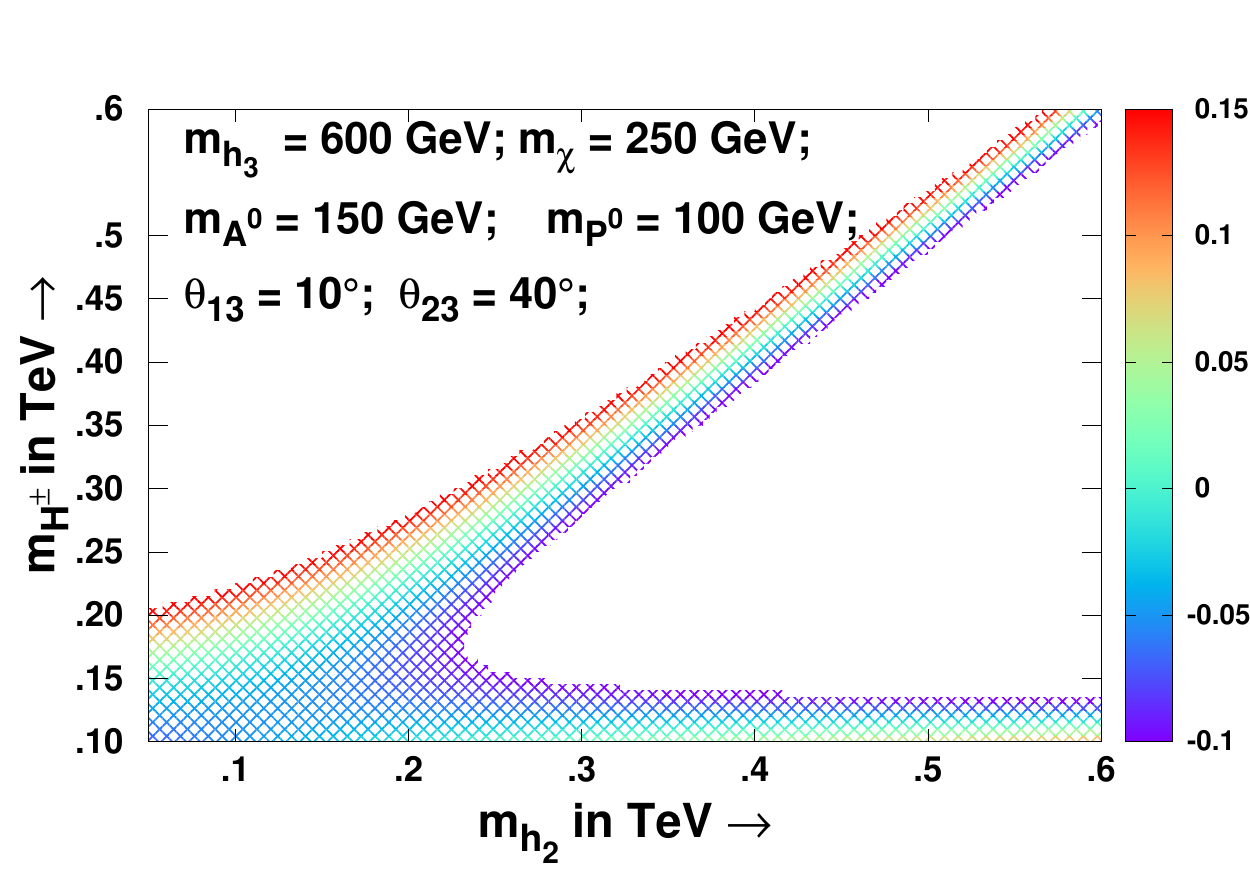}
\subcaption{\em{} }
\includegraphics[width=0.5\textwidth,clip]{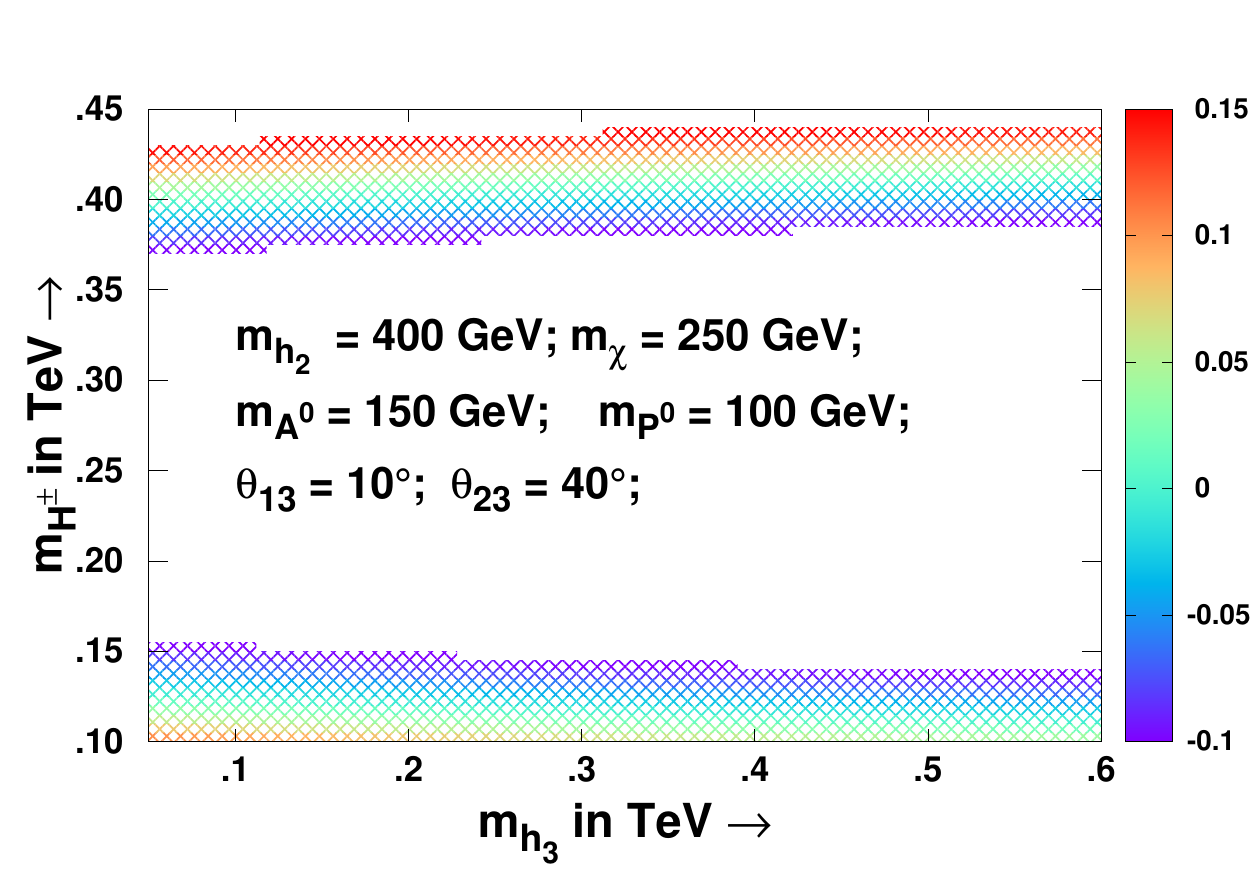}
\subcaption{\em{} }
\end{multicols}
\begin{multicols}{2}
\includegraphics[width=0.5\textwidth,clip]{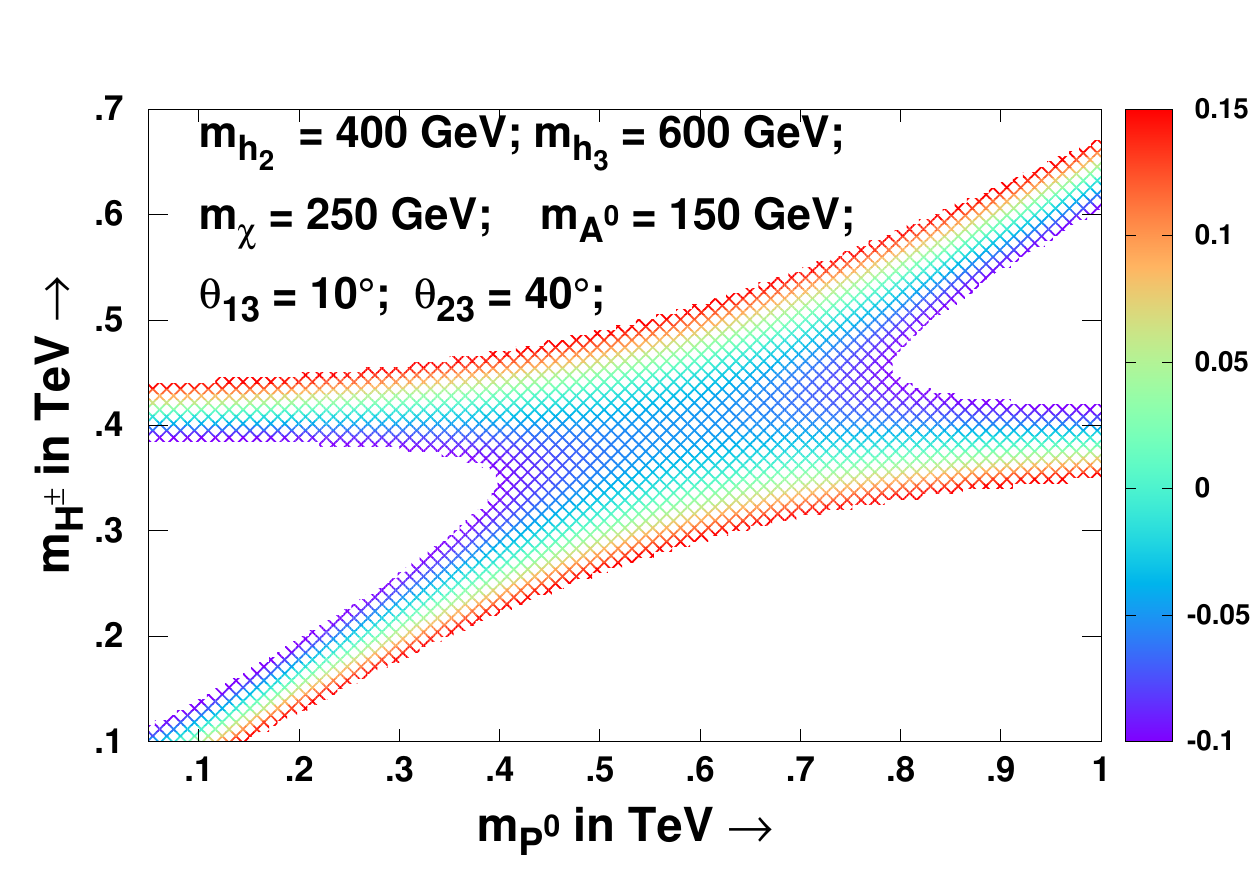}
\subcaption{\em{} }
\includegraphics[width=0.5\textwidth,clip]{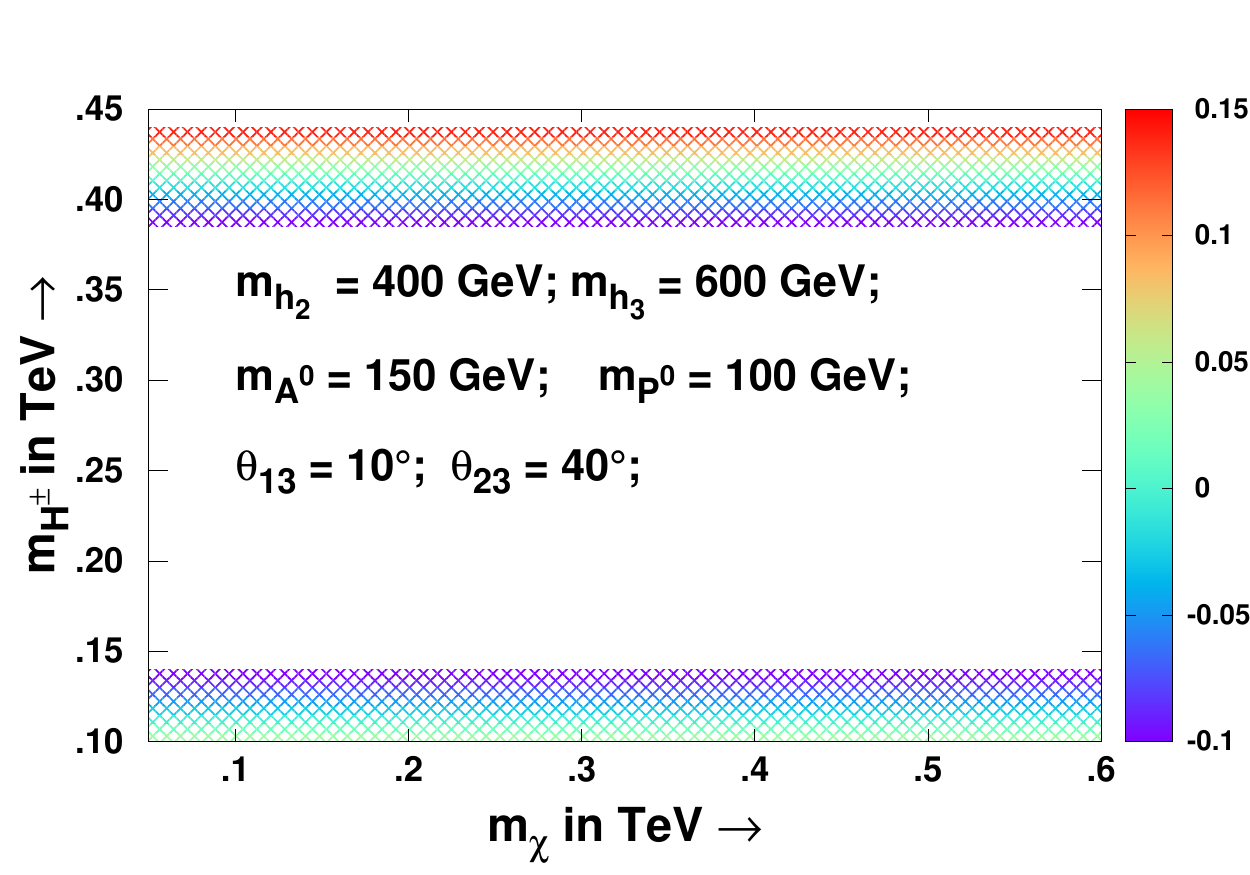}
\subcaption{\em{} }
\end{multicols}
\caption{ \em{One sigma shaded density maps in the plane defined by respective scalar/ pseudo-scalar/ VLL masses and charged Higgs mass $m_{H^\pm}$ satisfying the present experimental constraint on precision variable $\Delta T= 0.03 \pm 0.12$ \cite{PDG:2020}.}}
\label{STUconts}
\end{figure*}

In this subsection we compute the additional contribution to the precision observables from heavy physical mass eigenstates $m_{h_i,\, P^0,\,A^0,\,\chi}>> m_Z$. We constrain the scalar/ pseudo-scalar masses and their mixing angles based on the available  electroweak precision data.  Contributions to the oblique radiative corrections  are given in terms of  three precision parameters known as $S$, $T$ and $U$ \cite{Peskin:1991sw,Marciano:1990dp, Kennedy:1990ib, Kennedy:1991wa, Ellis:1992zi}. 
\par The precision observables derived from the radiative corrections of the gauge Boson propagator are essentially the two point vacuum polarization tensor functions of $\Pi^{\mu\nu}_{ij}\left(q^2\right)$  where $q$ is the four-momentum of the vector boson ($V=W,\ Z$ or $\gamma$). Following the prescription of the reference \cite{Haber:1993wf}  the  vacuum polarization tensor functions  corresponding to pair of gauge Bosons $V_i-V_j$ can be written as
\begin{subequations}
\bea
i\Pi^{\mu\nu}_{ij}(q)&=&ig^{\mu\nu}A_{ij}(q^2)+iq^\mu q^\nu
B_{ij}(q^2)\,\,\,\,{\rm where} \label{ewp1}\\
A_{ij}(q^2)&=&A_{ij}(0)+q^2F_{ij}(q^2) \label{ewp2}
\eea
\end{subequations}
where only $A_{ij}(q^2)$ are the relevant functions for the computation of precision observables. The equation \eqref{ewp2} defines the function $F_{ij}$.  Accordingly the precision parameters are defined as:
\begin{subequations}
\bea
S&\equiv& \frac{1}{g^2} \left(16\pi \cos\theta_{W}^2\right)\left[F_{ZZ}(m_Z^2)
-F_{\gamma\gamma}(m_Z^2)+\left({2\sin\theta_{W}^2-1\over \sin\theta_{W}
\cos\theta_{W}}\right)F_{Z\gamma}(m_Z^2)\right]\label{sdef}\\
T&\equiv & \frac{1}{\alpha_{em}}\left[{A_{WW}(0)\over m_W^2}-{A_{ZZ}(0)\over
 m_Z^2}\right] \label{tdef}\\
U&\equiv& \frac{1}{g^2}\left( 16\pi\right)\left[ F_{WW}(m_W^2)-F_{\gamma\gamma}(m_W^2)-
{\cos\theta_{W}\over \sin\theta_{W}}F_{Z\gamma}(m_W^2)\right]-S\,.\label{udef}
\eea
\end{subequations}
where $\alpha_{em}$ is the fine structure constant. It is worthwhile to mention that although $A_{ij}(0)$ and $F_{ij}$ are divergent by themselves but the total divergence associated with each precision parameter in equations \eqref{sdef},  \eqref{tdef} and \eqref{udef} vanish on taking into account a gauge invariant set of  one loop diagrams contributing  for a given pair of gauge Bosons.  

\par  The deviation from the predicted SM contribution for $S$ and $T$ parameters can be expressed analytically in terms of standard Veltman-Passarrino integrals $A_0,\, B_0$ and $B_{22}$  defined in the appendix \ref{app:VPfunctions}:
\begin{subequations}
\bea
\hspace*{-1cm} \Delta S&=& \frac{G_F\,\alpha_{em}^{-1}}{2\sqrt{2}\, \pi^2} \,\sin^2\left(2\,\theta_W\right)\,\Bigg[\sin^2\theta_{13} \bigg\{m_Z^2\bigg( \mathcal{B}_0(m_Z^2;m_Z^2,m_{h_1}^2)-  \mathcal{B}_0(m_Z^2;m_Z^2,m_{h_3}^2)   \bigg) \nn\\
&& +  \mathcal{B}_{22}(m_Z^2;m_Z^2,m_{h_3}^2) - \mathcal{B}_{22}(m_Z^2;m_Z^2,m_{h_1}^2)\bigg\}\nn\\
&&  + \sin^2\theta_{23} \bigg(\mathcal{B}_{22}(m_Z^2;m_{h_2}^2,m_{P^0}^2) - \mathcal{B}_{22}(m_Z^2;\mhpm^2,\mhpm^2) \bigg) \Bigg]
\eea 
 where
 \bea
\mathcal{B}_{22}(q^2;m_1^2,m_2^2) &\equiv& B_{22}(q^2;m_1^2,m_2^2)- B_{22}(0;m_1^2,m_2^2)\\
\mathcal{B}_{0}(q^2;m_1^2,m_2^2) &\equiv& B_{0}(q^2;m_1^2,m_2^2)- B_{0}(0;m_1^2,m_2^2)
\eea
\end{subequations}
 \bea
\Delta T&=& \frac{G_F\,\alpha_{em}^{-1}}{2\sqrt{2}\, \pi^2}\Bigg[ \sin^2\theta_{13} \Bigg\{m_W^2 \bigg(B_0(0;m_W^2,m_{h_1}^2)- B_0(0;m_W^2,m_{h_3}^2) \bigg)\nn\\
&& -m_Z^2 \bigg( B_0(0;m_Z^2,m_{h_1}^2) - B_0(0;m_Z^2,m_{h_3}^2) \bigg)+  B_{22}(0;m_W^2,m_{h_3}^2) - B_{22}(0;m_W^2,m_{h_1}^2)\nn\\
&&  +   B_{22}(0;m_Z^2,m_{h_1}^2)-  B_{22}(0;m_Z^2,m_{h_3}^2) \Bigg\}- \frac{1}{2}\, A_0(m^2_{H^\pm})+ B_{22}(0;\mhpm^2,m_{h_2}^2)\nn\\
&&+\cos^2\theta_{23} \bigg( B_{22}(0;\mhpm^2,m_{A^0}^2)-B_{22}(0;m_{h_2}^2,m_{A^0}^2)\bigg)   \nn\\
&&+ \sin^2\theta_{23} \bigg(B_{22}(0;\mhpm^2,m_{P^0}^2) - B_{22}(0;m_{h_2}^2,m_{P^0}^2)\bigg)  \nn\\
&& + \ 4\ \sin^4\theta_W \bigg(  m_\chi^2\ B_0(0;m_\chi^2,m_\chi^2)- 2 B_{22}(0;m_\chi^2,m_\chi^2) \bigg)  \Bigg]
\eea
\par The deviation in the theoretical predictions for precision observables in SM $\left( S_{\rm SM},\, T_{\rm SM}\right)$  from the electroweak precision measurements in the experiments for $U\ne 0$ $\left( S_{\rm expt},\, T_{\rm expt}\right)$ are parameterised as \cite{PDG:2020}  
\begin{subequations}
\bea
 \Delta S&=& S_{\rm expt.}- S_{\rm SM}=0.01 \pm 0.10\label{dscons}\\
 \Delta T&=& T_{\rm expt.}- T_{\rm SM}=0.03 \pm 0.12\label{dtcons}
 %\\ \Delta U&=& U_{\rm expt.}- U_{\rm SM}=0.02 \pm 0.11\label{ducons}
 \eea
\end{subequations}
\par Computing the precision observables numerically, we find that the allowed parameter space by the Higgs decays and LEP data satisfy the one sigma constraint for $\Delta S$ as given in equation \eqref{dscons}. In fact, the large uncertainty in the measurement allows the mass range between 50 - 1000 GeV for all scalars, pseudo-scalars and the VLL. However,  the constrain on $\Delta T$ as given in equation \eqref{dtcons} further shrinks  the allowed parameter region. We have depicted  the one sigma density maps for the allowed region by the $\Delta T$ in the plane defined by the masses  $m_{h_2}-m_{H^\pm}$,  $m_{h_3}-m_{H^\pm}$,  $m_{P^0}-m_{H^\pm}$ and $m_{\chi}-m_{H^\pm}$ in figure \ref{STUconts}. 
\par In the next section, we proceed with this constrained parameter space to look for viable explanation for the observed discrepancies in the measurements of the anomalous MDM for muon and electron.

%%%%%%%%%%%%%%%%%%%%%%%
\begin{figure}[h!]
\centering

\begin{multicols}{3}

\includegraphics[scale=0.2]{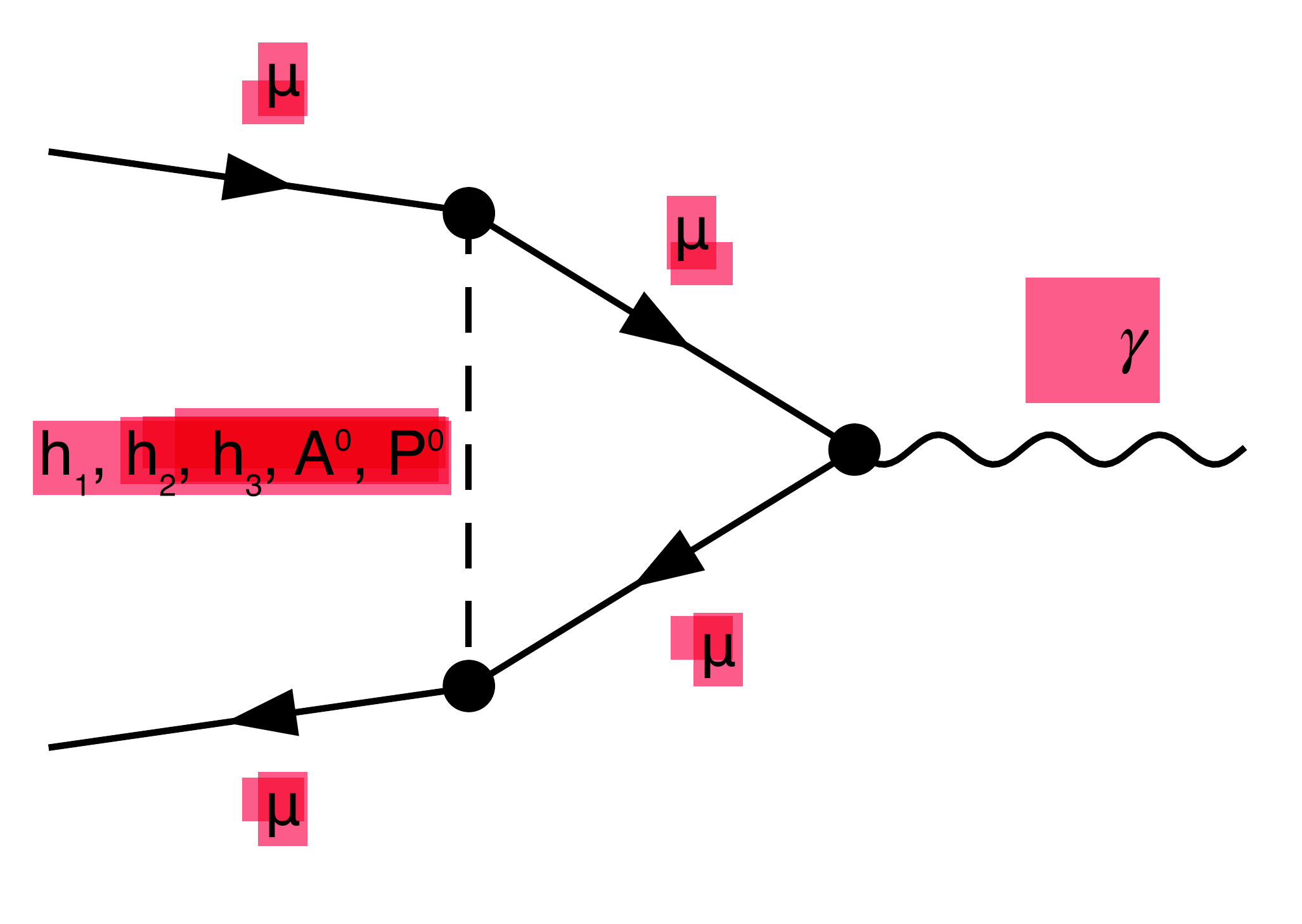}
\subcaption{Leptons}
\label{leponeloop}
\columnbreak
\includegraphics[scale=0.2]{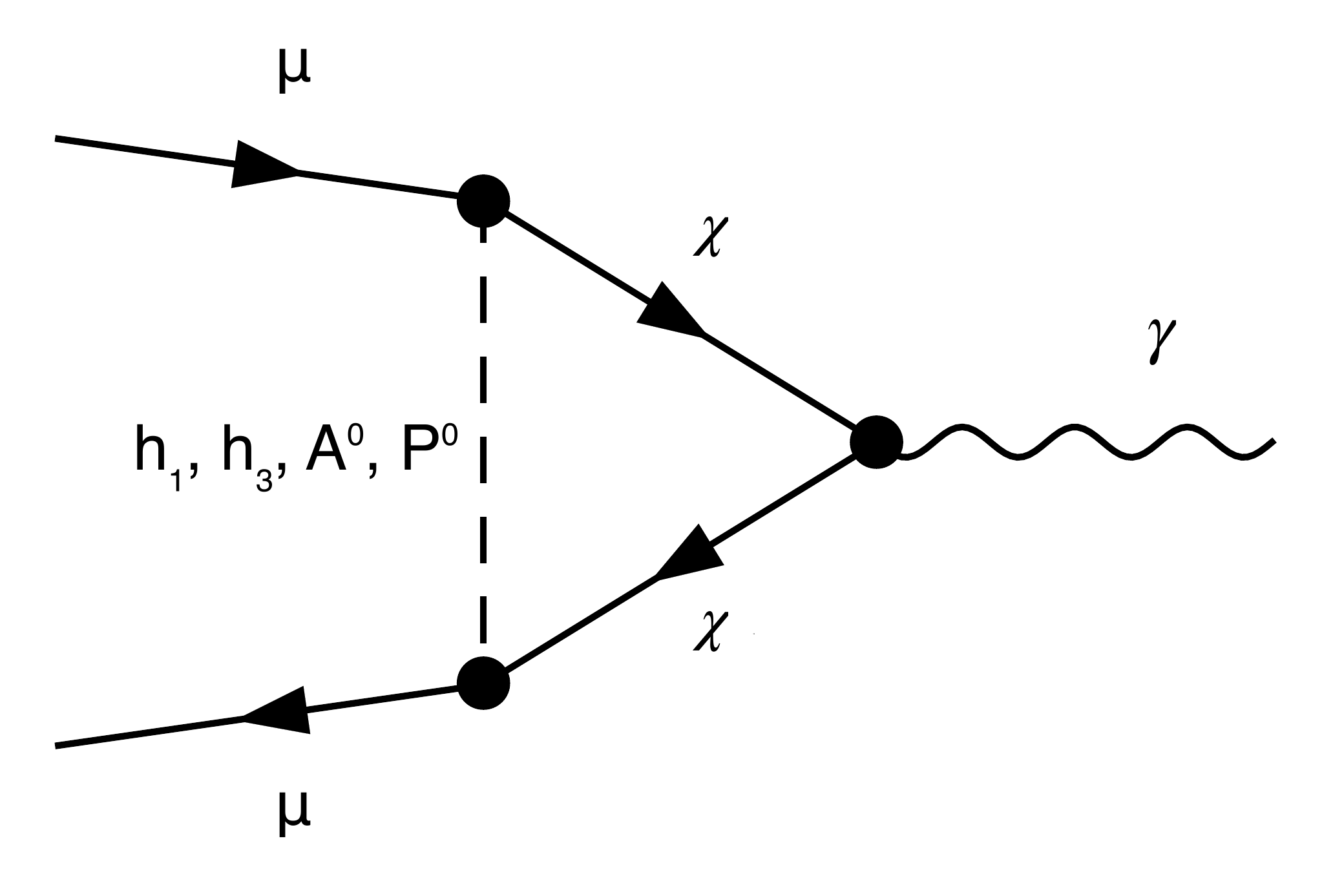}
\subcaption{\em{Vector-like Leptons}}
\label{VLoneloop}
\columnbreak
\includegraphics[scale=0.2]{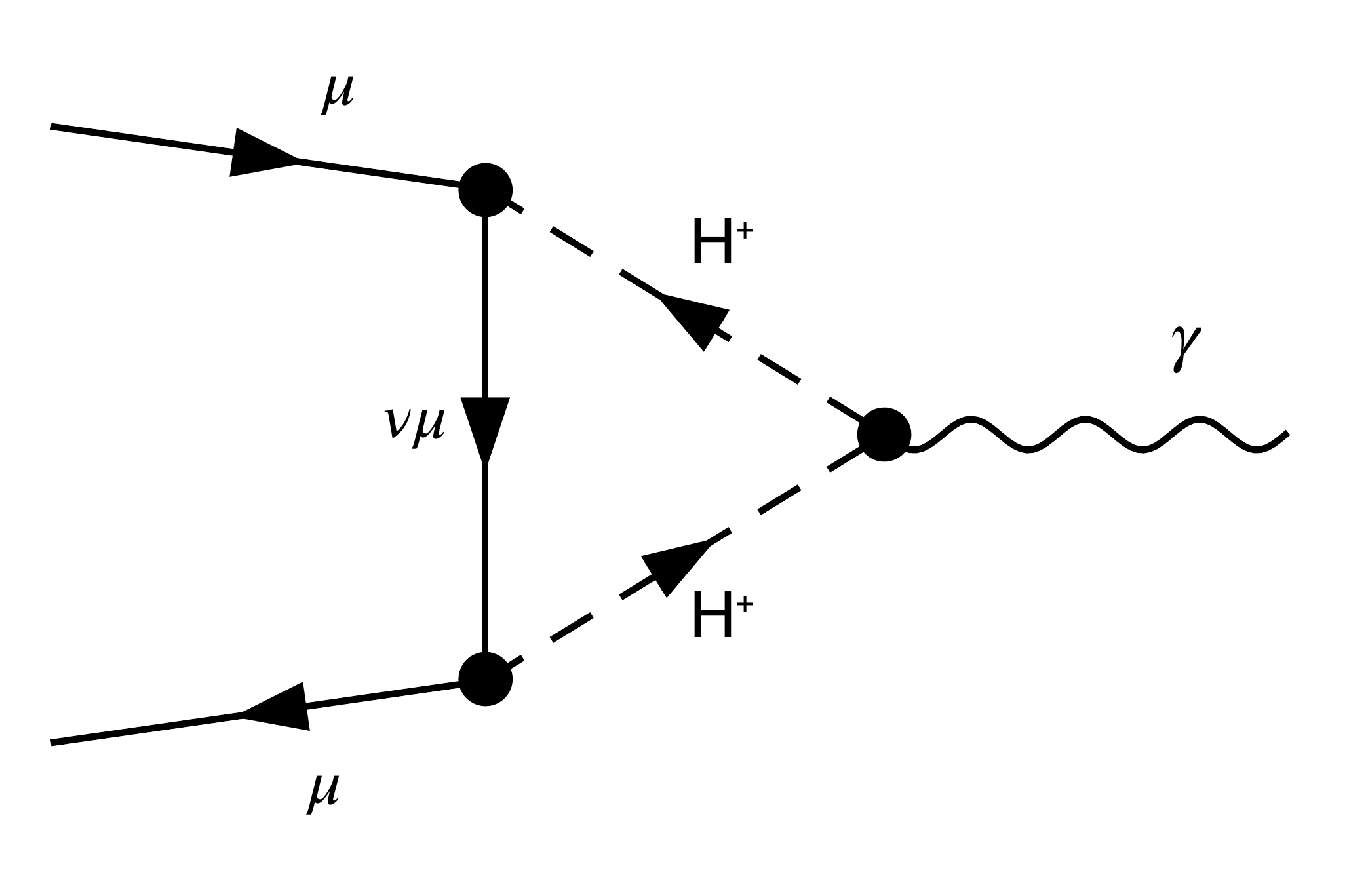}
\subcaption{\em{Charged Scalars}}
\label{ChargedscalOneloop}
\end{multicols}
\caption{\em{Dominant diagrams contributing to muon $g-2$ at one-loop level.}}
\label{OneLoop}
\end{figure}
\section{Anomalous MDM of Leptons}
\label{sec:MDM}
We compute the additional model contribution (other than SM diagrams)  to the $\Delta a_l$ at one and two loops respectively. 
\par In our model, the new one-loop contribution to $\Delta a_l$ arises from the exchange of CP-even and odd scalars and from the charged Higgs and VLL diagrams. We draw the additional (other than those allowed by SM)  dominant Feynman diagrams at one loop in figure \ref{OneLoop} based on the Lagrangian given in equations \eqref{scalarpot}, \eqref{SMYukawa} and \eqref{VLYukawa}. 
\par The sum of the contributions to lepton $\Delta a_l$  from the additional one-loop Feynman diagrams (other than SM) as shown in the figure \ref{OneLoop}  is calculated to be:
\begin{subequations} 
\bea
 \delta a_l^{\rm 1\,loop }&=& \frac{1}{16\, \pi^2}\left[ 2\ \frac{m_l^4}{v_{\rm SM}^2} \ \left( \frac{\cos^2\theta_{13}}{m_{h_1}^2}\  + \frac{\sin^2\theta_{13}}{m_{h_3}^2}- \frac{1}{m_{h^{\rm SM}}^2} \right)\ {\cal I}_1 +  m_l^2 \ \left( \frac{\cos^2\theta_{23}}{m_{A^0}^2} + \frac{\sin^2\theta_{23}}{m_{P^0}^2} \right)\  y_1^2\  {\cal I}_2 \right. \nn\\
&&  \left.  + \frac{m_l^2}{m_{h_2}^2}\ y_1^2\ {\cal I}_1 + \sum_{s_i=h_1, h_3, A^0, P^0} 
\left\vert y_{l\chi s_i}\right\vert^2\ \frac{m_l^2}{m_{s_i}^2}\ {\cal I}_3\ + \left\vert y_1\right\vert^2\ \frac{m_l^2}{m_{H^\pm}^2}\ {\cal I}_4 \right]
\label{eq:MDMoneloop}
\eea
\end{subequations}
\begin{figure}[htb]
\centering
\begin{multicols}{3}
\includegraphics[scale=0.2]{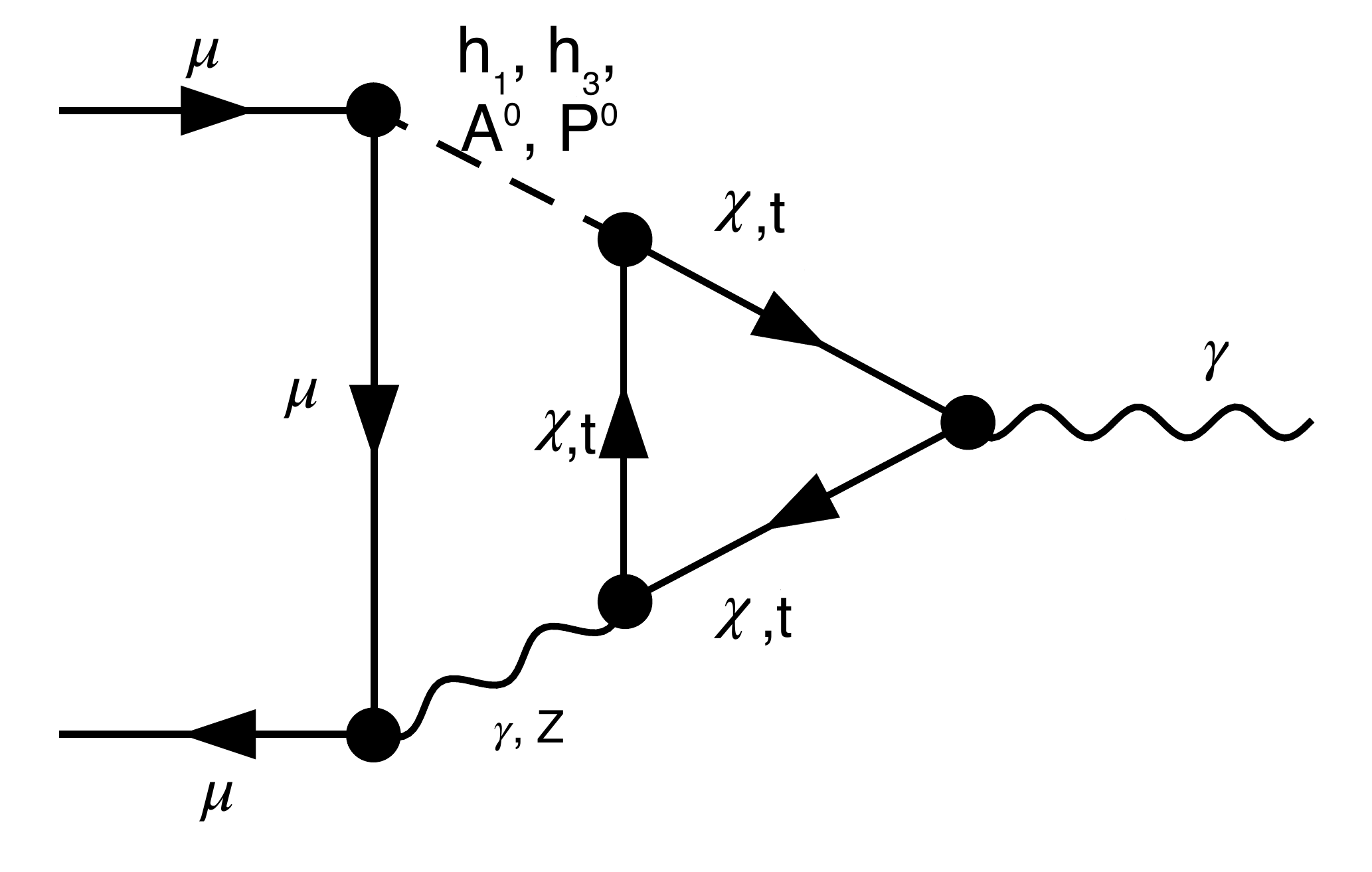}
\subcaption{Fermion triangle}
\label{VLtriangle}
\columnbreak
\includegraphics[scale=0.2]{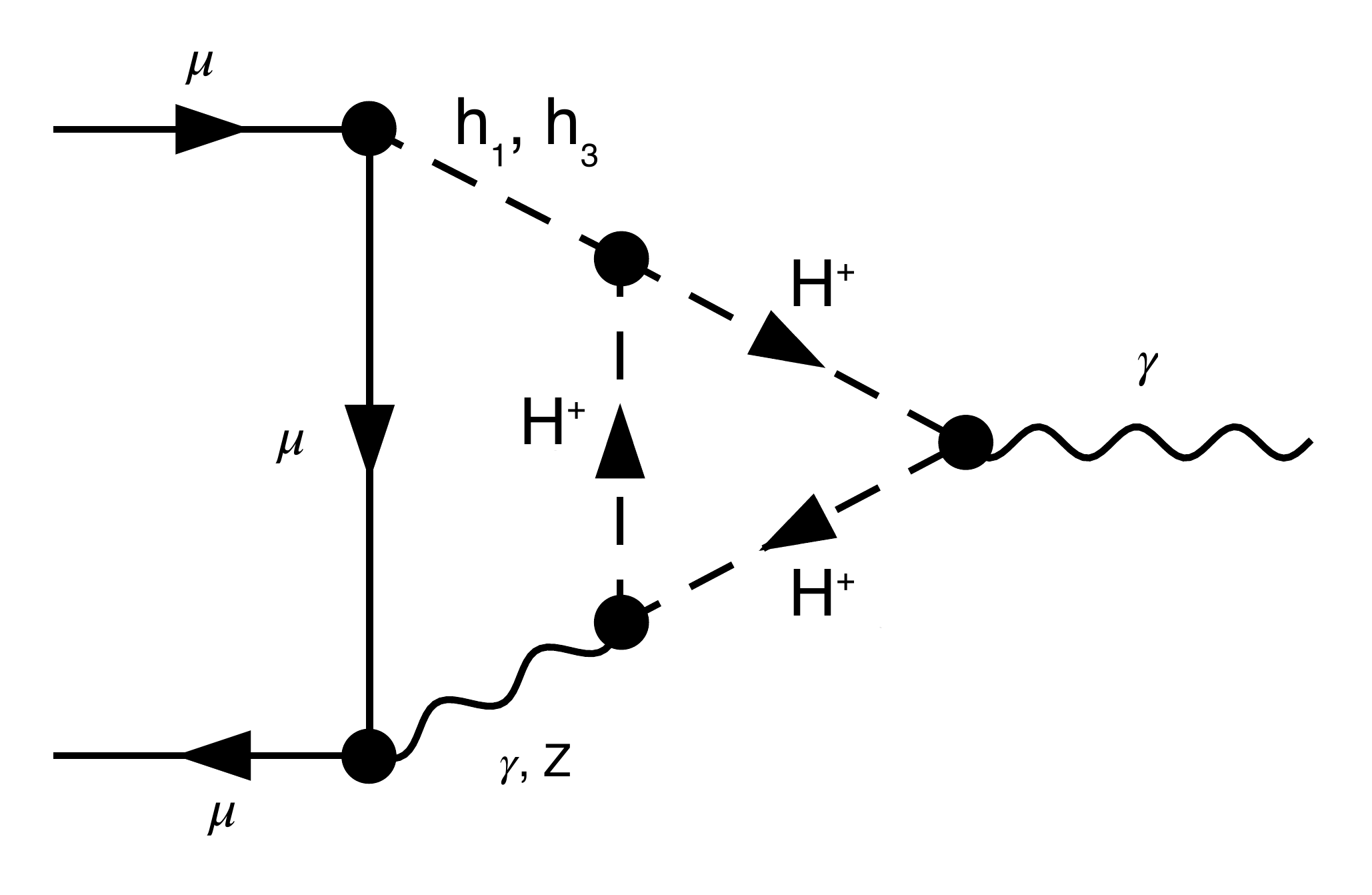}
\subcaption{$H\pm$ triangle}
\label{Hplustriangle}
\columnbreak
\includegraphics[scale=0.2]{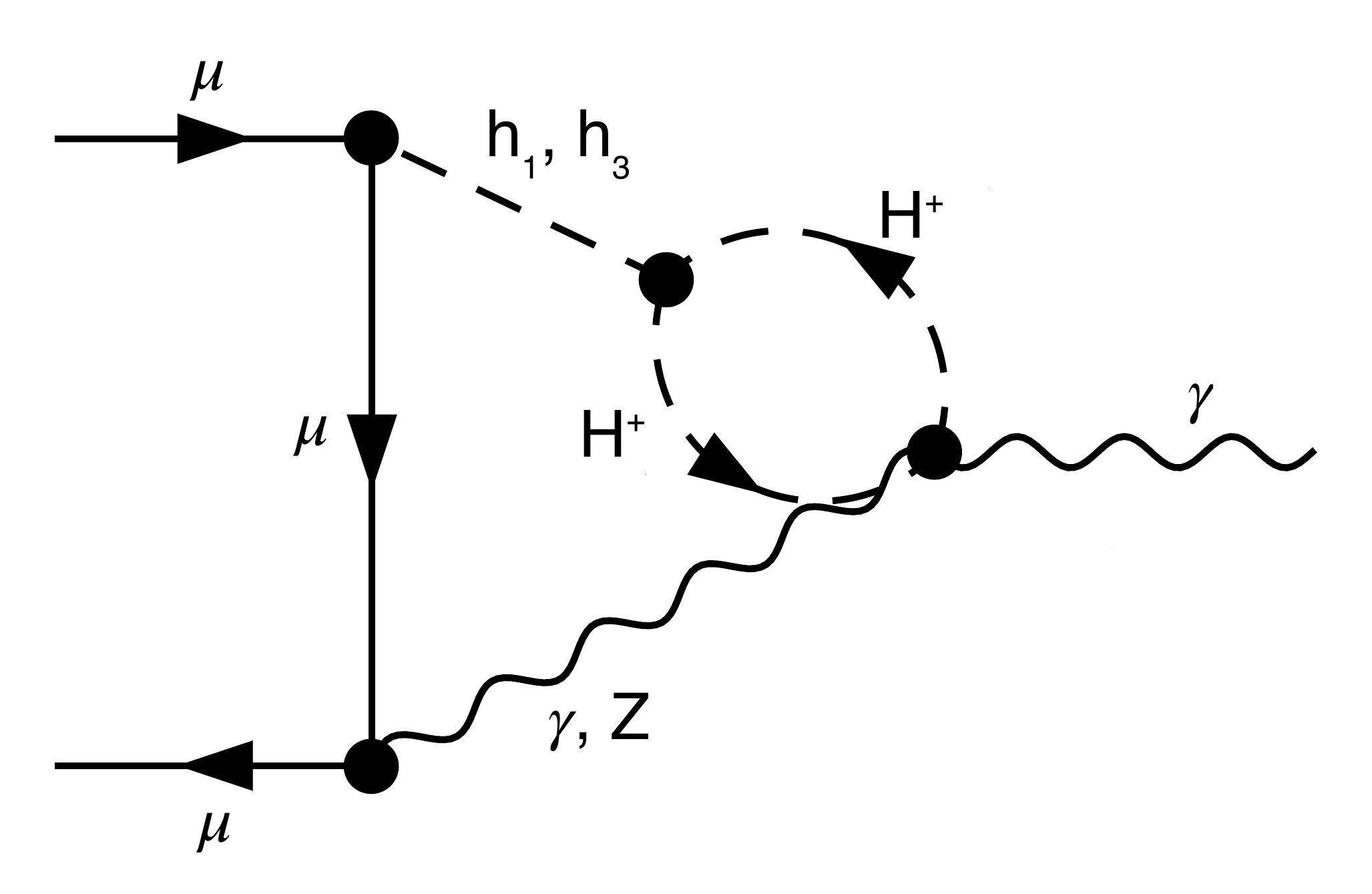}
\subcaption{$H^\pm$ bubble}
\label{Hplusbubble}
\end{multicols}
\caption{\em{Dominant diagrams contributing to muon $g-2$ at the two-loop level.}}
\label{TwoLoops:BarZee}
\end{figure}
where the one loop functions ${\cal I}_1, \,{\cal I}_2,\,{\cal I}_3$ and ${\cal I}_4$ are defined in the appendix \ref{app:MDMLoopFunc} in equations \eqref{i1eq},  \eqref{i2eq}, \eqref{i3eq} and \eqref{i4eq}   respectively. 
\par In order to obtain the common parameter-space satisfying both $\Delta a_\mu$ and $\Delta a_e$ which are of opposite signs, it is imperative to  analyse the nature of contribution by the mediating scalar/ pseudoscalar as given in figure \ref{OneLoop}. We observe that one-loop amplitudes in figure  \ref{leponeloop} are negative and positive corresponding to mediating pseudo-scalars and  scalars  respectively while loop amplitudes in figure \ref{VLoneloop}  are positive for both pseudo-scalars and  scalars except $h_2$ as it do not couples to VLL. The contribution from charged Higgs loop in figure  \ref{ChargedscalOneloop} is negative  and competitively much smaller in magnitude.
\begin{figure}[h!]
\centering
\begin{multicols}{2}
\includegraphics[width=0.5\textwidth,clip]{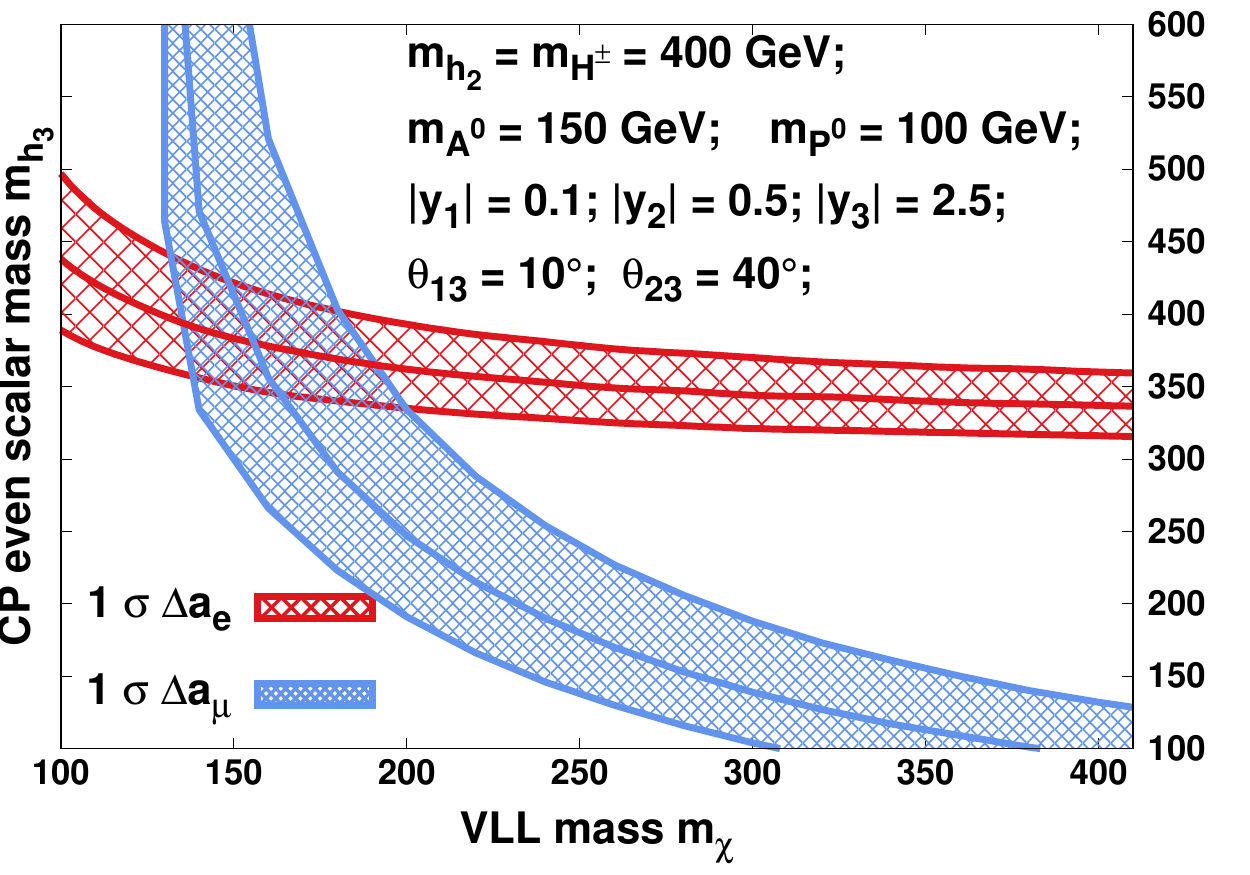}
\subcaption{\small \em{ }} \label{MDMcont1}
\includegraphics[width=0.5\textwidth,clip]{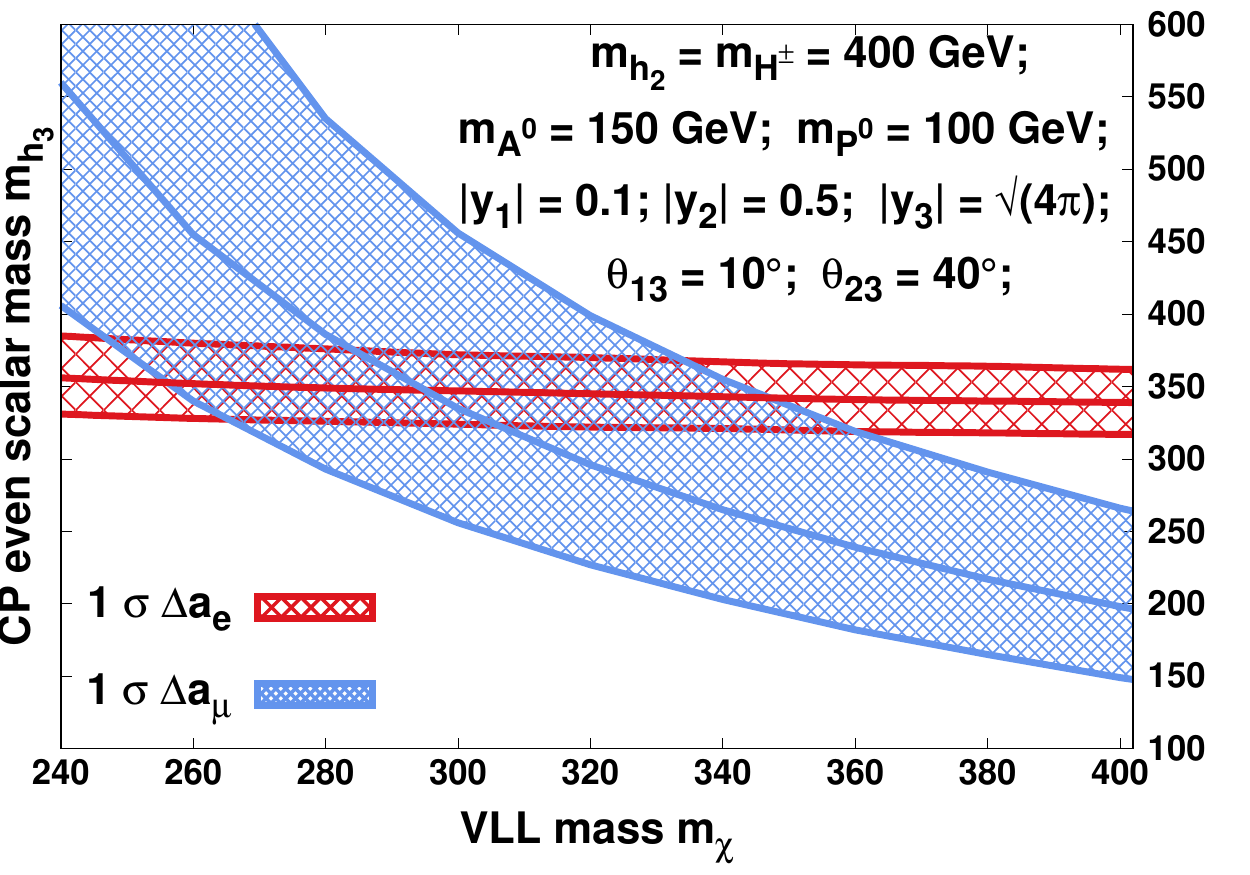}
\subcaption{\small \em{} }\label{MDMcont2}
\end{multicols}
\caption{ \em{Colored solid contours and associated one $sigma$ bands  are shown in $m_{h3}-m_\chi$ satisfying $\Delta a_{\mu} =  \left(251 \pm 59 \right) \times 10^{-11}$ \cite{Muong-2:2021ojo} (in blue) and $\Delta a_e = \left[-88\ \pm 28\, ({\rm expt.}) \pm 23\, (\alpha) \pm 2\,({\rm theory})\right] \times 10^{-14}$ \cite{Parker:2018vye} (in red). Contours are drawn for Yukawa couplings $\left\vert y_3\right\vert$ = 2.5 and the perturbative limit $\sqrt{4\pi}$ in figures \ref{MDMcont1} and \ref{MDMcont2} respectively, while other physical masses, couplings and angles remain same for the both. The overlap of the two bands depict the common parameter space.}}
\label{MDM1}
\end{figure}
%%%%
\begin{figure}[h!]
\centering
\begin{multicols}{2}
\includegraphics[width=0.5\textwidth,clip]{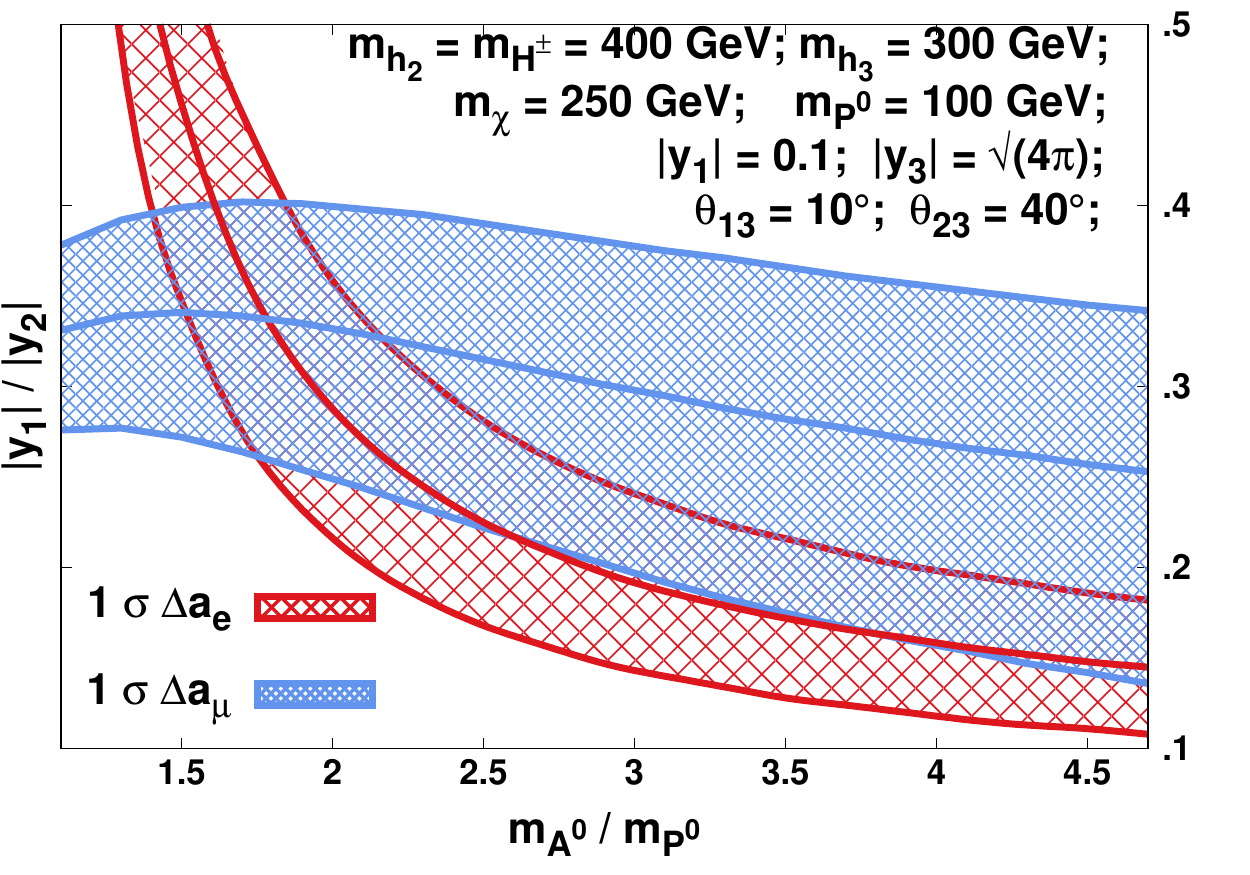}
\subcaption{\em{} }\label{MDMcont3}
\includegraphics[width=0.5\textwidth,clip]{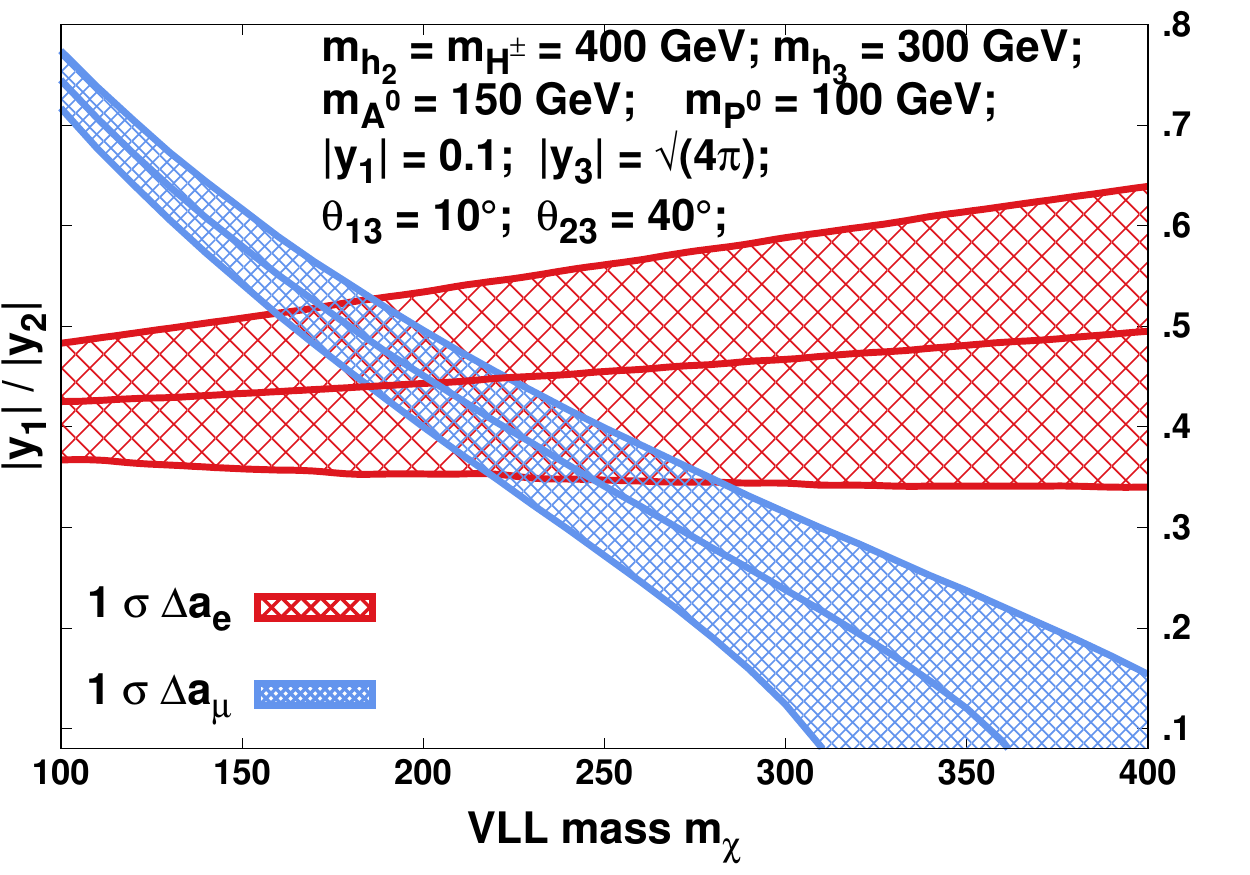}
\subcaption{\em{} }\label{MDMcont4}
\end{multicols}
\caption{ \em{Colored solid contours and associated one $sigma$ bands  are shown in   $\left\vert y_1\right\vert/\left\vert y_2\right\vert -  m_{A^0}/\,m_{P^0}$ and $\left\vert y_1\right\vert/\left\vert y_2\right\vert\,-\, m_\chi$ planes  satisfying  $\Delta a_{\mu} =  \left(251 \pm 59 \right) \times 10^{-11}$ \cite{Muong-2:2021ojo} (in blue) and $\Delta a_e = \left[-88\ \pm 28\, ({\rm expt.}) \pm 23\, (\alpha) \pm 2\,({\rm theory})\right] \times 10^{-14}$ \cite{Parker:2018vye} (in red).  Contours are drawn with fixed $m_\chi$ = 250  GeV and   $m_{A^0}$ = 150 GeV  in figures \ref{MDMcont3} and \ref{MDMcont4} respectively while other physical masses, couplings and angles remain same for the both. The overlap of the two bands depict the common parameter space. }}
\label{MDM2}
\end{figure}
\par  The contributions of two loop diagrams, some of which may dominate inspite of an additional loop suppression factor play a crucial role in the estimation of anomalous MDM. It is shown in the literature that the dominant two-loop Barr-Zee diagrams mediated by neutral scalars and pseudo-scalars can become relevant for certain mass scales so that their contribution to the muon anamalous MDM  are of the same order to that of one loop diagrams \cite{Chun:2020uzw}.
\par In figure \ref{TwoLoops:BarZee} we draw the dominant additional two-loop Barr-Zee diagrams (other than SM) contributing to the anomalous MDM of lepton which  are based on the Lagrangian given in equations \eqref{scalarpot}, \eqref{SMYukawa} and \eqref{VLYukawa}. The sum of the contributions to lepton $\Delta a_l$  from the additional two-loop Feynman diagrams (other than SM) as shown in the figure \ref{TwoLoops:BarZee}  is calculated to be:

 \begin{eqnarray}
\delta {a_l}^{2\ loop} &=&  \frac{\alpha_{em}}{4\ \pi^3}\ \left[ \frac{m_l}{v_{\rm SM}} \frac{y_2}{\sqrt{2}}\ \sin\theta_{13}\ \cos\theta_{13} \left\{ f\left( \frac{m_\chi^2}{m_{h_3}^2}\right) - f\left(\frac{m_\chi^2}{m_{h_1}^2}\right) \right\}\right.\nn\\
&&\left. + \frac{m_l}{v_{\rm SM}} \frac{m_t}{v_{\rm SM}}\ \left\{\sin^2\theta_{13}\ f\left( \frac{m_t^2}{m_{h_3}^2}\right) - \cos^2\theta_{13}\ f\left(\frac{m_t^2}{m_{h_1}^2}\right) +  f\left(\frac{m_t^2}{m_{h^{\rm SM}}^2}\right) \right\} \right.\nn\\
&&\left. + \frac{y_1\ y_2}{2}\ \frac{m_l}{m_\chi}\sin\theta_{23}\ \cos\theta_{23}\ \left\{ g\left( \frac{m_\chi^2}{m_{A^0}^2}\right)  - g\left(\frac{m_\chi^2}{m_{P^0}^2}\right) \right\}\right.\nn\\
&&\left. -\frac{ m_l^2}{4}\ \frac{m_l}{v_{\rm SM}^2}\left\{ \frac{ \cos\theta_{13}}{ m_{h_1}^2}\ g_{h_1 H^+ H^-}\ \tilde{f}\left( \frac{m_{H^\pm}^2}{m_{h_1}^2} \right) - \frac{ \sin\theta_{13}}{ m_{h_3}^2}\ g_{h_3 H^+ H^-}\ \tilde{f}\left( \frac{m_{H^\pm}^2}{m_{h_3}^2} \right) \right\} \right]\nn\\
\label{2loopal}
\end{eqnarray}
where the two loop functions $f$, $g$ and $\tilde f$ are defined in the appendix \ref{app:MDMLoopFunc} in equations \eqref{2loopint1}, \eqref{2loopint2} and \eqref{2loopint3} respectively.

\par We find that the two loop amplitudes with VLL triangle  in figure \ref{VLtriangle} are negative  for mediating scalars and positive for mediating pseudo-scalars. The two-loop contributions of the charged Higgs in figure \ref{Hplustriangle} and \ref{Hplusbubble} are comparatively small and negative. The dominant Barr-Zee contributions are found to depend on the mixing angle $\theta_{23}$, relative mass squared difference of the CP-odd scalars $m^2_{A^0}- m^2_{P^0}$ and the Yukawa couplings $y_1,\,y_2$. 
\par
\begin{figure}[h!]
\centering
\includegraphics[width=0.55\textwidth,clip]{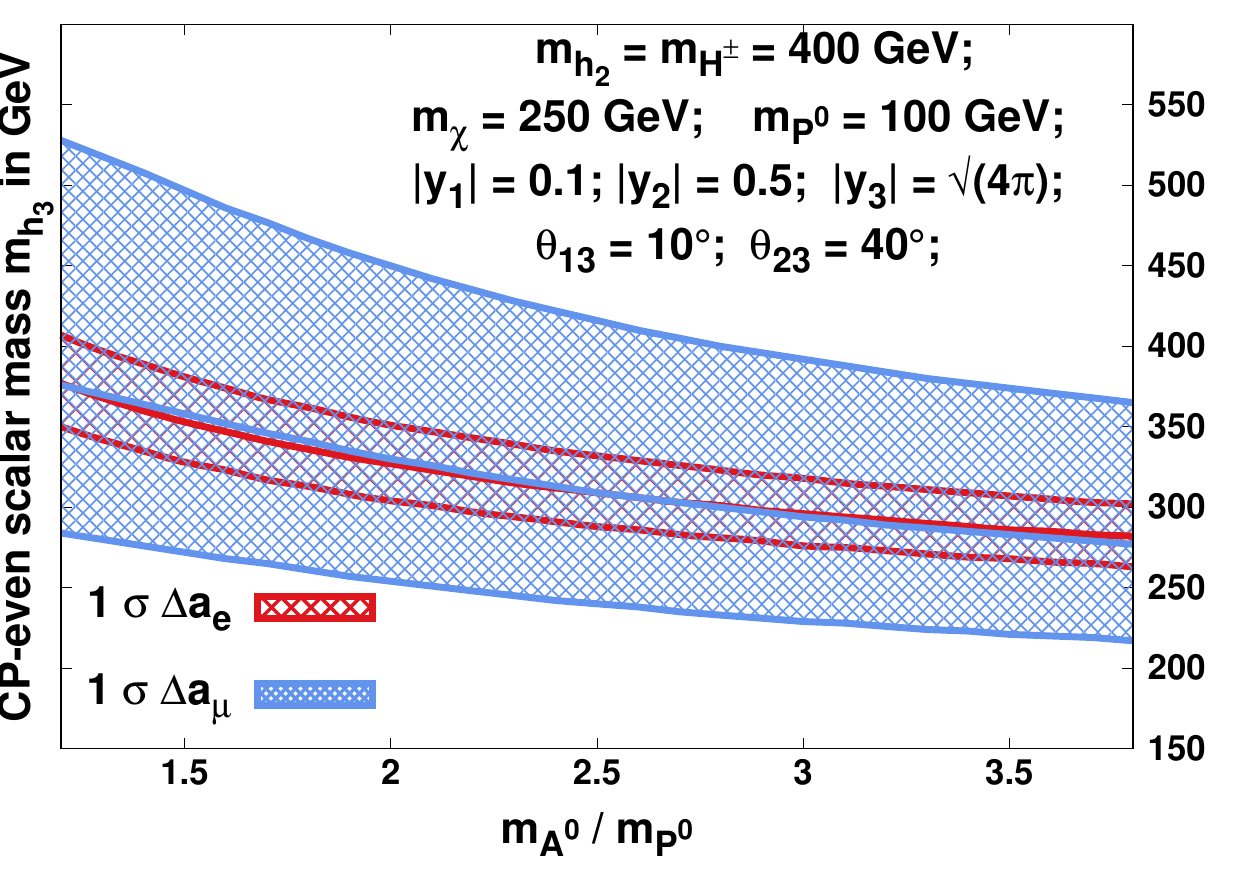}
\caption{ \em{Colored solid contours and associated one $sigma$ bands  are shown in $m_{h_3}- m_{A^0}/m_{P^0}$ plane satisfying  $\Delta a_{\mu} =  \left(251 \pm 59 \right) \times 10^{-11}$ \cite{Muong-2:2021ojo} (in blue) and $\Delta a_e = \left[-88\ \pm 28\, ({\rm expt.}) \pm 23\, (\alpha) \pm 2\,({\rm theory})\right] \times 10^{-14}$ \cite{Parker:2018vye} (in red). The overlap of the two bands depict the common parameter space.}} \label{MDMcont5}
\end{figure}

 The total contribution from one and two loop diagrams are computed for the constrained parameter space. Fixing  the physical masses $m_{H^\pm}\, =\,m_{h_2}$ = 400 GeV, Yukawa coupling $\left\vert y_1\right\vert$ = 0.1, $\Phi_1-\Phi_3$ mixing angle $\theta_{13}$ at 10$^\circ$ and $\Phi_2-\Phi_3$ mixing angle $\theta_{23}$ at 40$^\circ$  for our analysis, we vary  $m_{h_3},\, m_{\chi},\, m_{A^0}$, $\left\vert y_2\right\vert$  and $\left\vert y_3\right\vert$ to account for the contribution to anomalous MDM as observed in the experiments. Contours   fulfilling the central value for $\Delta a_\mu$ \cite{Muong-2:2021ojo} and $\Delta a_e$ \cite{Parker:2018vye} quoted  in equations \eqref{delamu} and \eqref{delae} respectively    are depicted in $m_\chi - m_{h_3}$ plane as shown in figure \ref{MDMcont1} for $\left\vert y_3\right\vert$ = 2.5 and in figure \ref{MDMcont2} for the maximum allowed value of $\left\vert y_3\right\vert$ = $\sqrt{4\pi}$ respectively. The overlapping region of the allowed one sigma  bands for $\Delta a_\mu$ in blue  and $\Delta a_e$ in red  exhibit the common  parameter space satisfying both the experimental  observations simultaneously. It is observed that overlapping region broadens for the large Yukawa coupling $\left\vert y_3\right\vert$ corresponding to  a narrow mass range for $m_{h_3}$.
In figure  \ref{MDMcont3} and \ref{MDMcont4}  we fix $m_{h_3}$ = 300 GeV, $\left\vert y_3\right\vert =\sqrt{4\pi}$ and depict the   one sigma band of contours  in $\left\vert y_1\right\vert/\left\vert y_2\right\vert -  m_{A^0}/\,m_{P^0}$  plane for $m_\chi$ = 250 GeV  and $\left\vert y_1\right\vert/\left\vert y_2\right\vert - m_{\chi}$ plane for $m_{A^0}$ = 150 GeV respectively. We observe that a common parameter space for both are possible for $m_{A^0}\ge 1.5\, m_{P^0} $ and 170 GeV $\lesssim m_\chi\lesssim$ 300 GeV.   Choosing VLL mass at 250 GeV, $\left\vert y_3\right\vert =\sqrt{4\pi}$  and  pseudo-scalar mass range such that $m_{A^0} > m_{P^0}$,  we plot  the narrow red one sigma band for $\Delta a_e$  contours in $m_{h_3}-\left(m_{A^0}/\,m_{P^0}\right)$ plane in figure \ref{MDMcont5} and observe that it completely overlaps with the broad blue one sigma band for $\Delta a_\mu $ contours. This constrains  the $m_{h_3}$ to be $<$ 350 GeV.

%%%%
\section{Summary}
\label{sec:summary}
In this article we considered an extended inert 2HDM model with an SM singlet complex scalar and a singlet vector like lepton field to explain the observed anomalies in the muon and electron dipole magnetic moments. The model parameters are expressed in terms of the physical masses and mixing angles of the CP even and odd scalars and are constrained from the Higgs decay to a pair of gauge Bosons at LHC, LEP data and electro-weak precision measurements. The contribution of scalars and vector-like lepton arises at the dominant one-loop and two-loop Barr-Zee diagrams. The CP even scalar  one-loop contributions to $\Delta a_l$ are positive whereas the contribution from the CP-odd scalars is negative. The contribution of the VLL is important and decreases with the VLL mass. The Barr-Zee contributions mainly depend on the mixing angle $\theta_{23}$ and on relative mass squared difference of CP-odd scalars $A^0$ and $P^0$ and decreases with the VLL mass.   

\par The constrained model is systematically analysed to accommodate both the experimental observations simultaneously. We depict the viable common parameter space through  overlapping one sigma band of contours satisfying $\Delta a_\mu$ \cite{Muong-2:2021ojo} and $\Delta a_e$ \cite{Parker:2018vye} simultaneously in figures  \ref{MDMcont1}, \ref{MDMcont2}, \ref{MDMcont3}, \ref{MDMcont4} and \ref{MDMcont5}. We find that there exists a fairly large common  parameter space where the anomalous magnetic moments of muon and electron can be explained.

%%%%%%%%%%%%%%%%%%%%%%%%%%%%%%%%%%%%%%%%%%%%
\acknowledgments
We acknowledge the partial financial support from SERB grant CRG/2018/004889. HB acknowledges the CSIR JRF fellowship. 						
\appendix
\begin{center}
{\bf \large Appendix}
\end{center}
\section{ Scalar Couplings in terms of Mass Parameters}
\label{App:InputParams}

\begin{enumerate}
\item Since, LHC data favours a SM-like Higgs $\sim$ 125 GeV, we  allign the mass of the lightest neutral Higgs state $m_{h_1}$ coming predominantly from the doublet $\Phi_1$ with the SM Higgs $m_{h^{\rm SM}}$. In order to accomodate the $2\,\sigma$  uncertainty of the measured Higgs mass \cite{PDG:2020} the variation for $0.2\le\l_1\le 0.3$  may be allowed through mixing of $\Phi_1 - {\Phi_3}$.

\item The quartic parameter $\lambda_2$ appears only in the quartic interaction of $Z_2$-
odd particles and is therefore not constrained by our analysis.

 \item Considering  VEVs $v_{\rm SM}$ and $v_{s}$, mixing angles $\theta_{13}$ and $\theta_{23}$, coupling $\lambda_{13}$ and  masses 
$m_{22}^2,\,\, m_{h_1}^2, \,\, m_{h_2}^2, \,\, m_{h_3}^2, \,\, m_{H^\pm}^2, \,\, m_{A^0}^2, \,\, {\rm and} \,\, m_{P^0}^2$ to be the free parameters, we can express
$m_{11}^2,\,\, m_{33}^2,\,\, \lambda_3,\,\, \lambda_4,\,\, \lambda_5 \,\, \lambda_8, \,\,\lambda_{11} \,\, {\rm and} \,\, \kappa$ in terms of the above free parameters.

\item The $Z_2$-odd charged scalar $H^\pm$ comes 
solely from the second doublet, as in the IDM; its mass is given by 
\begin{equation}
m_{H^\pm}^2 = \frac{1}{2} \left[- m_{22}^2 + v_{\rm SM}^2 \lambda_3 +  v_{s}^2\, \lambda_{13}\right]
\end{equation}
$\lambda_3$ can be expressed in terms of free parameters $\lambda_{13}$, $m_{22}^2 $ and the charged Higgs mass as
\begin{eqnarray}
\lambda_3= \frac{1}{v_{\rm SM}^2} \left[2\,m^2_{H^\pm} + m^2_{22} -\lambda_{13}\,\, v_s^2\right]
\end{eqnarray}
Notice, that the mass relations for the $Z_2$-odd sector from the IDM  hold, namely 
\begin{eqnarray}
m_{h_2}^2 &=& m_{H^\pm}^2 + \frac{v_{\rm SM}^2}{2} \left[\lambda_4 + \lambda_5\right],\\
 m_{A^0}^2 + m_{P^0}^2 &=& m_{H^\pm}^2 + 
\frac{v_{\rm SM}^2}{2} \left[\lambda_4 - \lambda_5\right] \label{relIDM}
\end{eqnarray}
Therefore
\begin{eqnarray}
\lambda_4 &=& \frac{1}{v_{\rm SM}^2} \left[m_{h_2}^2+ m_{A^0}^2+ m_{P^0}^2-2\,m_{H^\pm}^2\right] \label{lam4}
\end{eqnarray}

\item From pseudoscalars and heavy neutral scalar masses we have 
\begin{equation}
\lambda_5 = \frac{1}{v_{\rm SM}^2}\left[m_{h_2}^2-m_{A^0}^2- m_{P^0}^2\right] \label{lam5}
\end{equation}

\item From mass relations for $m_{h^0}$ and $m_{S^0}$ given in equations (3.19) and (3.20) we get
\begin{eqnarray}
\lambda_8 &=& \frac{1}{v_{s}^2}\left[m_{h_1}^2 + m_{h_3}^2 -\lambda_1\, v_{\rm SM}^2\right]\\
\lambda_{11} &=& \frac{1}{v_{\rm SM}\,v_{s}}\left(\lambda_1\, v_{\rm SM}^2 -\lambda_8\, v_{s}^2\right)\tan\left(2\,\theta_{13}\right)
\end{eqnarray}

 \item The heavy neutral scalar mass from the Inert doublet is given as
\begin{eqnarray} 
 m_{h_2}^2 = \frac{1}{2} \left[ -m_{22}^2 + v_{\rm SM}^2 \lambda_{345}+ v_{s}^2\, \lambda_{13}\right] 
\end{eqnarray}
Therefore
\bea
\lambda_{13}&=& \frac{1}{v_{s}^2}\left(2\  m_{h_2}^2+ m_{22}^2- v_{\rm SM}^2 \lambda_{345}\right)
\eea
 \end{enumerate} 
 \section{Definition of Loop Form Factors}
 \label{app:formfactors}
 
 The loop amplitudes are expressed in terms  of dimensionless parameters $\tau$ and $\lambda$, which are essentially function of the masses of physical scalars/ pseudo-scalars and fermions. 
 
 \begin{subequations}
\begin{eqnarray}
{\cal M}^{\gamma\gamma}_0(\tau)&=&-\tau[1-\tau f(\tau)]\\
{\cal M}^{\gamma\gamma}_{1/2}(\tau)&=&2\tau[1+(1-\tau)f(\tau)],\\
{\cal M}^{\gamma\gamma}_1(\tau)&=&-[2+3\tau+3\tau(2-\tau)f(\tau)]\\
{\cal M}^{Z\gamma}_{0} (\tau,\lambda) & = & I^{Z\gamma}_1(\tau,\lambda)\\
{\cal M}^{Z\gamma}_{1/2} (\tau,\lambda) & = & \left[I^{Z\gamma}_1(\tau,\lambda) - I^{Z\gamma}_2(\tau,\lambda)
\right] \\
{\cal M}^{Z\gamma}_1 (\tau,\lambda) & = & c_W \left\{ 4\left(3-\frac{s_W^2}{c_W^2} \right)
I^{Z\gamma}_2(\tau,\lambda) + \left[ \left(1+\frac{2}{\tau}\right) \frac{s_W^2}{c_W^2}
- \left(5+\frac{2}{\tau} \right) \right] I_1^{Z\gamma}(\tau,\lambda) \right\} \nn\\
\label{eq:hzgaform}
\end{eqnarray}
\end{subequations}
with
\bea
f(\tau)=
\left\{ 
\begin{array}{cc}
\arcsin^2\big(\frac{1}{ \sqrt{\tau} }\big) & \textrm{for } \, \tau\geqslant 1,\\
-\frac{1}{4}\Big[\log\Big(\frac{1+\sqrt{1-\tau}}{1-\sqrt{1-\tau}}\Big)-i\pi\Big]^2  & \textrm{for } \,\tau<1.\\
\end{array}
\right.
\eea
The functions $I^{Z\gamma}_1$ and $I^{Z\gamma}_2$ 
are given by
\begin{eqnarray}
I^{Z\gamma}_1(\tau,\lambda) & = & \frac{\tau\lambda}{2(\tau-\lambda)}
+ \frac{\tau^2\lambda^2}{2(\tau-\lambda)^2} \left[ f(\tau)-f(\lambda) 
\right] + \frac{\tau^2\lambda}{(\tau-\lambda)^2} \left[ g\left(\frac{1}{\tau} \right) - 
g\left(\frac{1}{\lambda}\right) \right] \nn \\
I^{Z\gamma}_2(\tau,\lambda) & = & - \frac{\tau\lambda}{2(\tau-\lambda)}\left[ f(\tau
)- f(\lambda) \right]
\end{eqnarray}
Function $g(\tau)$ can be expressed as
\begin{equation}
g(\tau) = \left\{ \begin{array}{ll}
\displaystyle \sqrt{\tau^{-1}-1} \arcsin \sqrt{\tau} & \ {\rm for}\ \tau \geqslant 1 \\
\displaystyle \frac{\sqrt{1-\tau^{-1}}}{2} \left[ \log \frac{1+\sqrt{1-\tau
^{-1}}}{1-\sqrt{1-\tau^{-1}}} - i\pi \right] &\ {\rm for}\ \tau  < 1
\end{array} \right.
\label{eq:gtau}
\end{equation}

\section{Veltman Passarino Loop Integrals}
\label{app:VPfunctions}
The $A_0,\, B_0,\, B_{22}$ integrals are defined as 
\begin{subequations}
\bea
A_0(m^2)&=& m^2\ \left( \Delta + 1 - \ln m^2  \right),\\
B_0(q^2;m_1^2,m_2^2)&=&\Delta-\int_0^1\,dx\,\ln(X-i\epsilon)\\
B_{22}(q^2;m_1^2,m_2^2)&=&\frac{1}{4}(\Delta+1)\left[m_1^2+m_2^2
-\frac{1}{3} q^2\right]-\frac{1}{2}\int_0^1\,dx\,X\ln(X-i\epsilon)
\eea
\end{subequations}
where $X\equiv m_1^2x + m_2^2(1-x) - q^2 x(1-x)$ and $\Delta\equiv{2\over 4-d}+\ln(4\pi)+\gamma_E$ in $d$ space-time dimensions.

\section{One loop and two loop functions for MDM}
\label{app:MDMLoopFunc}
The one loop functions ${\cal I}_1,\, {\cal I}_2,\, {\cal I}_3$ and ${\cal I}_4$ required to compute one loop contribution to the MDM of leptons \eqref{eq:MDMoneloop} are defined as 
\begin{subequations}
\bea
{\cal I}_1(r^2)&=&\int_0^1 dx\ \frac{(1+x)(1-x)^2}{(1-x)^2\ r^2+ x}\label{i1eq}\\
 {\cal I}_2(r^2)&=&\int_0^1 dx\ \frac{-(1-x)^3}{(1-x)^2\ r^2+ x},\label{i2eq}\\
{\cal I}_3(r'^2)&=& \int_0^1 dx\ \frac{x(1-x)^2}{(1-x)\ r'^2+ x}\label{i3eq}\\ \hskip 3cm {\cal I}_4(r^2)&=& \int_0^1 dx\ \frac{-x(1-x)}{1- (1-x)\  r^2}\label{i4eq}
\eea
\end{subequations}
with $r=\frac{m_l}{m_{s_i}},\ r'= \frac{m_\chi}{m_{s_i}} $ and $s_i= h_1,\ h_2,\ h_3,\ A^0,\  P^0.$

\par The two-loop functions $f(r^2 )$, $g(r^2 )$  and $\tilde{f}(r^2 )$ contributing to the MDM of leptons given in equation \eqref{2loopal}  are defined as
\begin{subequations}
\begin{eqnarray}
f(r^2) &=& \frac{r^2}{2}\, \int_0^{1} dx\ \frac{1\ -\ 2x(1-x)}{x(1-x)\ -\ r^2}\
\ln \left[\,\frac{x(1-x)}{r^2}\,\right]\label{2loopint1}\\
g(r^2) &=& \frac{r^2}{2}\, \int_0^{1} dx\ \frac{1}{x(1-x)\ -\ r^2}\
\ln \left[\,\frac{x(1-x)}{r^2}\,\right] \label{2loopint2}\\
\tilde{f}(r^2) &=& \int_0^{1} dx\ \frac{x(1-x)}{r^2 -\ x(1-x)}\
\ln \left[\,\frac{x(1-x)}{r^2}\,\right] \label{2loopint3}
\end{eqnarray}
\end{subequations}
%%%%%%%%%%%%%%%%%%%%%%%%%%%%%%%%%%%%%%%%%%%%%%%%%%

\end{document}